\documentclass[aps,prb,twocolumn,10pt,longbibliography,floatfix,superscriptaddress,nofootinbib]{revtex4-2}
\usepackage{physics}
\usepackage{bbold}
\usepackage[english]{babel}
\usepackage[utf8x]{inputenc}
\usepackage{amssymb}
\usepackage{amsmath}
\usepackage{dsfont}
\usepackage{comment}
\usepackage{lipsum}  
\usepackage[normalem]{ulem}
\usepackage[pdftex]{graphicx}
\usepackage{xspace}
\usepackage{dsfont}
\usepackage{bm}
\usepackage[export]{adjustbox}
\usepackage[pdfencoding=auto, psdextra]{hyperref}
\usepackage{cleveref}
\hypersetup{
    colorlinks,%
    citecolor=blue,%
    filecolor=blue,%
    linkcolor=blue,%
    urlcolor=blue
}
\usepackage{tabularx}

\usepackage[usenames, dvipsnames]{xcolor}

\providecommand{\kp}[0]{\texorpdfstring{$\bm{k}\cdot\bm{p}$}{kp}\xspace}

\newcommand{\E}{\bar{E}}

\begin{document}

\title{Dual topological insulator with mirror symmetry protected helical edge states}

\author{Warlley H. Campos}
\email{Corresponding author: warlleyhcampos@gmail.com}
\affiliation{Instituto de F\'{i}sica de S\~{a}o Carlos, Universidade de S\~{a}o Paulo, 13560-970 S\~{a}o Carlos, S\~{a}o Paulo, Brazil}
\affiliation{Institut für Physik, Johannes Gutenberg-Universität Mainz, 55099 Mainz, Germany}
\author{Poliana H. Penteado}
\affiliation{Instituto de F\'{i}sica de S\~{a}o Carlos, Universidade de S\~{a}o Paulo, 13560-970 S\~{a}o Carlos, S\~{a}o Paulo, Brazil}
\affiliation{Department of Physics, University of Basel, Klingelbergstrasse 82, CH-4056 Basel, Switzerland}
\author{Julian Zanon}
\affiliation{Department of Applied Physics and Science Education, Eindhoven University of Technology, 5612 Eindhoven AZ, Netherlands}
\author{Paulo E. \surname{Faria~Junior}}
\affiliation{Institut f\"{u}r Theoretische Physik, Universit\"{a}t Regensburg, D-93040 Regensburg, Germany}
\author{Denis R. Candido}
\affiliation{Department of Physics and Astronomy, University of Iowa, Iowa City, Iowa 52242, USA}
\author{J. Carlos Egues}
\affiliation{Instituto de F\'{i}sica de S\~{a}o Carlos, Universidade de S\~{a}o Paulo, 13560-970 S\~{a}o Carlos, S\~{a}o Paulo, Brazil}
\affiliation{Department of Physics, University of Basel, Klingelbergstrasse 82, CH-4056 Basel, Switzerland}

\begin{abstract}
Dual topological insulators (DTIs) are simultaneously protected by time-reversal and crystal symmetries, representing advantageous alternatives to conventional topological insulators. By combining \emph{ab initio} calculations and the \kp approach, here, we investigate the electronic band structure of a Na$_2$CdSn triatomic layer and derive a low-energy $4\times 4$ effective model consistent with all the symmetries of this material class. We obtain the effective Hamiltonian using the Löwdin perturbation theory, the folding-down technique, and the theory of invariants and  determine its  parameters by fitting our analytical dispersion relations to those of \emph{ab initio} calculations. We then calculate the bulk topological invariants of the system and show that the Na$_2$CdSn triatomic layer is a giant-gap  (hundreds of millielectronvolts) quasi-two-dimensional DTI characterized by both spin and mirror Chern numbers $-2$. In agreement with the bulk-boundary correspondence theorem, we find that  a finite-width strip of Na$_2$CdSn possesses two pairs of counterpropagating helical edge states per interface. We obtain analytical expressions for the edge state energy dispersions and wave functions, which are shown to agree with our numerical calculations. Our work opens an avenue for further studies of Na$_2$CdSn as a potential DTI candidate with room-temperature  applications in areas of technological interest, such as nanoelectronics and spintronics.
\end{abstract}

\maketitle

\section{Introduction}\label{sec:intro}

Since the theoretical proposal and experimental observation of two-dimensional (2D) \cite{kane2005z2,kane2005quantum,fu2006time,bernevig2006quantumprl,bernevig2006quantum,konig2007quantum} and three-dimensional (3D) \cite{Fu2007a,Fu2007,qi2008topological,Zhang2009,liu2010model,Hughes2011} topological insulators (TIs), these materials have been the subject of substantial research in condensed matter physics and related areas of science \cite{liu2010model,hasan2010colloquium,qi2011topological,Campos2017,han2018quantum,Campos2018a}. Further investigations soon resulted in the discovery of another important class of topological materials named topological crystalline insulators (TCIs) \cite{Fu2011,liu2014spin,Ando2015,hsu2016two,hsu2019topology}. Both TIs and TCIs host midgap boundary states related to a bulk topological invariant via the bulk-boundary correspondence \cite{qi2006general,mong2011edge}. The TI phase can be characterized by  a nontrivial spin Chern number (in 2D) \cite{sheng2005nondissipative,sheng2006quantum,fukui2007topological,prodan2009robustness} or $Z_2$ invariant \cite{kane2005z2,fu2006time,Fu2007}, whereas the characterization of a TCI phase depends on the material and involves at least one crystal symmetry, e.g.,  inversion, rotation, or the mirror operation [it may or may not involve time-reversal symmetry (TRS)]. Well-known examples are TCIs characterized by the mirror Chern number \cite{Hsieh2012,Liu2013,Ando2015,zhou2018topological} and the $Z_2$ invariant defined in terms of the product of time reversal with a $C_4$ rotation \cite{Fu2011}.

The interplay between time-reversal and crystalline symmetries on the topology of 3D materials was investigated in Bi$_{1-x}$Sb$_x$ ~\cite{teo2008surface,matsuda2017surface} and Bi$_2$Te$_3$~\cite{rauch2014dual} alloys,  which were predicted to be not only $Z_2$ TIs but also TCIs characterized by a nontrivial mirror Chern number. Such a dual topological character defines Bi$_{1-x}$Sb$_x$, Bi$_2$Te$_3$, and other materials \cite{eschbach2017bi1te1,lee2022robust,facio2019dual,ghosh2019saddle,marrazzo2020emergent} as dual TIs (DTIs), considered more robust against external perturbations than each constituent phase \cite{li2023two,lee2022robust,rauch2014dual}. For example, it has been shown that, to open a gap in the Dirac cone describing the $(111)$ surface states of Bi$_2$Te$_3$,  one must break time-reversal and mirror symmetries simultaneously \cite{rauch2014dual}.

DTIs in 2D have been much less studied than their 3D counterparts. The literature is nearly limited to theoretical studies of the Na$_3$Bi single-layer system~\cite{niu2017robust,acosta2019spin,Focassio2020}. More recently, the rectangular Bi bilayer~\cite{li2023two}  and the hexagonal IrO \cite{li2023twoa} have also been proposed. Up to now, to the best of our knowledge, these have not yet been experimentally realized. 

In this paper, we explore an overlooked DTI constituted by a triatomic layer of the ternary sodium compounds Na$_2XY$ ($X=$ Mg, Cd; $Y=$ Pb, Sn), which can potentially be obtained by mechanical exfoliation of 3D van der Waals Dirac semimetals \cite{peng2018} already synthesized experimentally \cite{schuster1980ternare,yamada2014synthesis}. We derive a \kp Hamiltonian for this class of materials (provided that the ordering of the energy bands around the Fermi level is the same as that for Na$_2$CdSn) and analyze the bulk topology and the corresponding mirror symmetry protected helical edge states of its Na$_2$CdSn representative. By carrying out density functional theory (DFT) calculations, we first obtain the electronic band structure of Na$_2$CdSn, including the irreducible representations (irreps), orbital composition, and spin texture of the energy bands (Fig. \ref{fig:fig1}).  Then we derive our $4\times 4$ model Hamiltonian for the Na$_2XY$ material class by employing group theory analysis and three techniques well established in the \kp literature, namely, the L{\"o}wdin perturbation theory \cite{lowdin1951anote,winkler2003book}, the folding-down approach~\cite{bernardes2007, Calsaverini2008intersubband, fu2020spinorbit}, and the theory of invariants \cite{bir1974book,winkler2003book}.

To obtain the parameters of our model, we diagonalize our effective Hamiltonian and fit its analytical energy dispersions to our DFT results. We then investigate the orbital (pseudospin) composition and spin texture of the effective band structure (Fig. \ref{fig:fig2}). As a consequence of mirror symmetry \cite{kurpas2019spin} (see Appendix \ref{sec:mirrorsymmetry}), the energy bands are spin polarized along the $z$ direction, in agreement with the DFT calculations (Fig. \ref{fig:fig1}). Furthermore, by regularizing our model and numerically evaluating the Chern number \cite{bernevig2013book} for each spin sector, we show that Na$_2$CdSn is a DTI with the TI and TCI phases characterized by spin and mirror Chern numbers $-2$,  respectively (Fig. \ref{fig:regplots}).

We investigate the DTI spectrum for two finite systems: a semi-infinite plane (vacuum/DTI) and a ribbon geometry (vacuum/DTI/vacuum). For the semi-infinite plane, we employ the envelope function approach to analytically derive two pairs of counterpropagating topological edge states related by TRS (Kramers pairs) whose dispersion relations depend quadratically on the crystal momentum (Fig. \ref{fig:edgestates_fitted_params}). For the ribbon geometry, we use the finite difference method to numerically obtain its electronic band structure and identify the existence of four pairs of counterpropagating edge states, two pairs located on each interface of the system (Fig. \ref{fig:ribbonresults}). The states localized at opposite sides of the ribbon are quasidegenerate, with a small spin splitting caused by the bulk inversion asymmetry (BIA). Our results for both geometries are in agreement with the bulk-boundary correspondence theorem \cite{qi2006general,mong2011edge}.

The extra robustness against external perturbations  \cite{li2023two,lee2022robust,rauch2014dual} positions DTIs as promising materials for applications in nanotechnology. The Na$_2$CdSn system investigated here is even more compelling. According to our DFT calculations, its expected bulk energy gap, $E_g \approx 234.8\,$meV, is $\sim$1 order of magnitude larger than those of 2D TIs based on HgTe/CdTe, InAs/GaSb or InAs$_{0.85}$Bi$_{0.15}$/AlSb quantum wells, for which $E_g$ $\lesssim 30\,$meV~\cite{konig2007quantum,Knez2011,knez2012quantum,Candido2018}. Such features showcase this ternary compound as an outstanding candidate for the study of room-temperature topological effects and the development of nanoelectronic \cite{xue2011topological,strunz2020interacting,cao2023future}, spintronic \cite{awschalom2009spintronics,pesin2012spintronics,mellnik2014spin}, thermoelectric \cite{gresta2019optimal,badura2024observation,surgers2024anomalous}, and optical \cite{du2017evidence, syperek2022observation} devices.

We note that a study of the Na$_2XY$ triatomic layers was performed in Ref. \cite{mao2019dual}. The authors propose an effective model Hamiltonian \footnote{The authors of Ref. \cite{mao2019dual} provide very limited information about the basis
set used in their work. This precluded us from verifying whether their reported effective Hamiltonian respects all the symmetries of the system -- in particular time-reversal symmetry.} that can be immediately seen to differ from ours by (i) the lack of terms proportional to $(k_x\pm i k_y)^2$ in the off-diagonal matrix elements and (ii) the presence of terms proportional to $k_x\pm i k_y$ in the secondary diagonal, where $\bm{k}=(k_x,k_y)$ represents the crystal momentum. It is important to emphasize that we consistently kept terms up to second order in $\bm{k}$ in the perturbation theory. 

As we discuss in Sec. \ref{sec:topologyanalysis}, the terms proportional to $(k_x\pm i k_y)^2$ play a crucial role in the calculation of the  spin and mirror Chern numbers, which are both equal to $-2$. Moreover, they  give rise to the quadratic energy dispersions  of the corresponding topological edge states. The topological invariants obtained from our analytical model differ significantly from the spin and mirror Chern numbers $-1$ reported in Ref. \cite{mao2019dual}. Following  the bulk-boundary correspondence theorem, we predict the existence of two pairs of helical edge states at each boundary of a finite Na$_2XY$ system, whereas Ref. \cite{mao2019dual} reports only a single pair.

To ensure the accuracy of our model, we first performed the derivation of the \kp Hamiltonian using the L{\"o}wdin perturbation theory, and separately, using the folding-down approach. We also derived our effective Hamiltonian using the theory of invariants.  The results obtained are in complete agreement with each other and differ from the model proposed in Ref. \cite{mao2019dual}.  In addition, our investigation here goes beyond that of Ref. \cite{mao2019dual} by providing effective \kp parameters for the Na$_2$CdSn DTI and deriving analytical expressions for both the energy dispersion and wave functions of the edge states arising in this material.

This paper is organized as follows. In Sec. \ref{sec:DFTresults}, we present the crystal structure and DFT calculations for Na$_2$CdSn triatomic layers. Section \ref{sec:kpmodel} is devoted to the \kp model obtained via the L{\"o}wdin perturbation theory. We also present our results regarding the fitting of the effective model to the energy bands obtained via DFT. In Sec. \ref{sec:topologyanalysis}, we analyze the topology of Na$_2$CdSn, and in Sec. \ref{sec:edgestates}, we calculate the analytical dispersion relations for the topological edge states. We also present our numerical study of the topological edge states in a ribbon geometry. We summarize our findings in Sec. \ref{sec:conclusions}.

\section{N\MakeLowercase{a}$_2$C\MakeLowercase{d}S\MakeLowercase{n} triatomic layer: Crystal structure and DFT results}
\label{sec:DFTresults}

\begin{figure*}[t]
    \centering
    \includegraphics[width=.98\linewidth]{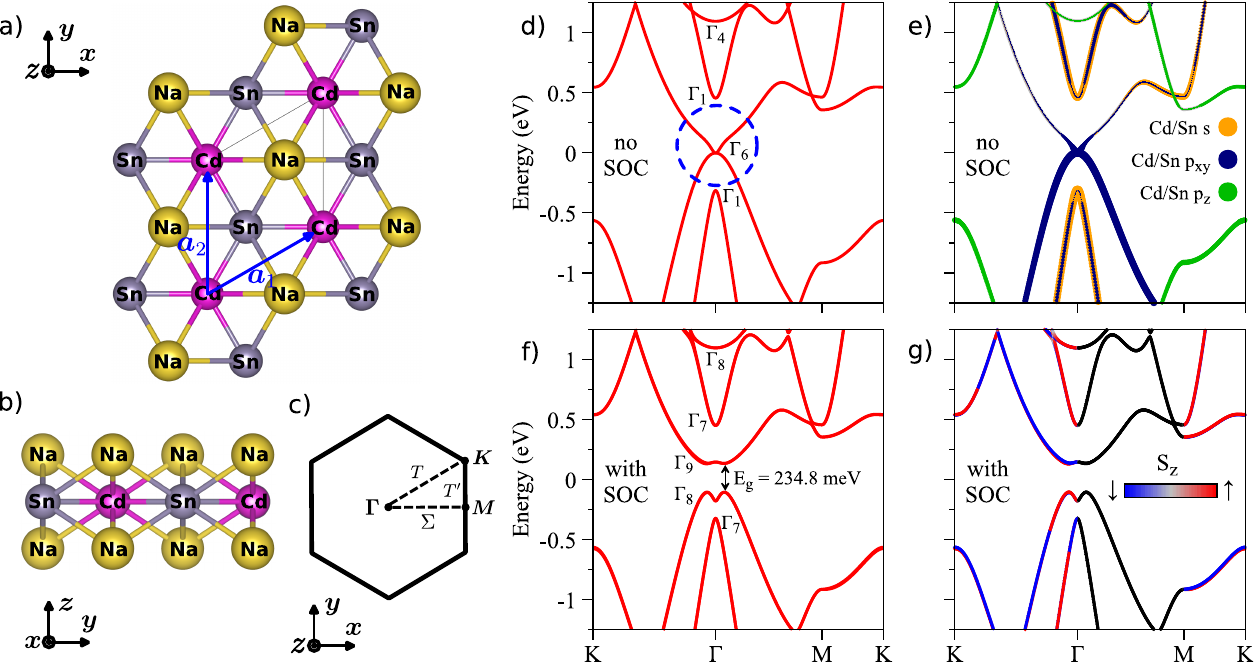}
    \caption{(Color online) a) Top  and b) side views of the quasi-two-dimensional (2D) honeycomb crystal structure of Na$_2$CdSn. The Cd and Sn atoms crystallize in a hexagonal lattice, while Na atoms lie above and below the center of each hexagon. The coordinate system used is shown according to the respective view. c) Brillouin zone (BZ) for the quasi-2D honeycomb crystal structure. $\Gamma=(0,0)$, $M=\frac{2\pi}{\sqrt{3}a}(1,0)$, and $K=\frac{2\pi}{3a}(\sqrt{3},1)$, where $a$ is the lattice constant. The $\Gamma-M$, $\Gamma-K$, and $M-K$ high-symmetry lines are labeled by $\Sigma$, $T$, and $T'$, respectively. d) Electronic band structure of Na$_2$CdSn without spin-orbit coupling (SOC). The two $\Gamma_6$ bands are degenerate at the Fermi level ($E_F=0$), characterizing a semimetal. The region of interest, to be described by the effective model, is indicated by the blue dashed circle. e) Orbital composition of the band structure without SOC. The $\Gamma_6$ bands (close to $E_F$) are dominated by the $p_x$ and $p_y$ orbitals (dark blue) of Cd and Sn, while the $\Gamma_1$ bands are dominated by the $s$ orbital. f) When SOC is considered, a gap of $E_g \,=\, 234.8\,$meV opens up at the Fermi level, and we can see that the material is a semiconductor. Furthermore, the energy bands are spin degenerate along the $\Gamma-M$ high-symmetry line but present a very small spin splitting along the other directions. g) Spin texture of the electronic band structure with SOC. The energy bands have spin expectation value of $\langle S_z\rangle \approx \pm \hbar/2$ [$\langle S_z\rangle>0$ ($\langle S_z\rangle<0$) red (blue)], while $\langle S_x\rangle=\langle S_y\rangle= 0$. The doubly degenerate bands along $\Gamma-M$ are colored black, for $\langle S_z \rangle$ is not univocally defined along this direction.}
    \label{fig:fig1}
\end{figure*}

The Na$_2$CdSn triatomic layer is constituted by Cd and Sn atoms forming a coplanar hexagonal lattice, while the sodium (Na) atoms lie above and below the center of each hexagon, forming a quasi-2D honeycomb material (see Fig. \ref{fig:fig1}).  This quasi-2D structure can be found as the constituting stacking block of the 3D Na$_2$CdSn Dirac semimetal, already synthesized experimentally \cite{schuster1980ternare,yamada2014synthesis}. In recent first-principles calculations, Mao \emph{et al.} \cite{mao2019dual} have verified the stability of these compounds by performing phonon spectrum analysis and revealing only positive phonon energies.

In Fig. \ref{fig:fig1}-a), we show a top view of the Na$_2$CdSn. The coordinate system is chosen such that the $z$ axis points out of the plane of the figure.  The primitive lattice vectors read $\bm{a}_1=\frac{a}{2}\left(\sqrt{3},1\right)$ and $\bm{a}_2=\left(0,a\right)$, where $a$ is the lattice parameter. In Fig. \ref{fig:fig1}-b), we show a side view of the crystal structure with the observer looking from the positive $x$ axis. From this angle, the $x$ axis points out of the plane of the figure, and we can see more clearly the position of the Na atoms relative to Cd and Sn. The reciprocal lattice vectors can be shown to be $\bm{b}_1=\left(\frac{4\pi}{\sqrt{3}a},0\right)$ and $\bm{b}_2=\frac{2\pi}{\sqrt{3}a}\left(-1,\sqrt{3}\right)$. The first Brillouin zone (BZ) in this coordinate system is shown in Fig. \ref{fig:fig1}-c), along with the high-symmetry points  $\Gamma=\left(0,0\right)$, $M=\left(\frac{2\pi}{\sqrt{3}a},0\right)$ and $K=\frac{2\pi}{3a}\left(\sqrt{3},1\right)$ \cite{CastroNeto2009}.

The crystal structure of the 3D bulk compound belongs to the nonsymmorphic space group $D_{6h}^4$ ($P6_3/mmc$), with point group $D_{6h}$ ($6/mmm$) \cite{yamada2012synthesis,yamada2014synthesis,peng2018,Wang2012,liu2014discovery}. Our DFT results show that the Na$_2$CdSn triatomic layer belongs to the $D_{3h}^1$ ($P\bar{6}m2$) space group, whose crystallographic point group is $D_{3h}$ ($\bar{6}m2$) \cite{yamada2012synthesis,yamada2014synthesis}. %
The corresponding set of generators are the $C_3$ rotation about the $z$ axis ($C_{3z}$), $C_2$ rotation about the $x$ axis ($C_{2x}$), and the mirror plane reflection about the $xy$ plane ($\sigma_h$) [see character table in the Supplemental Material (SM) \cite{Na2CdSnsupmat} (see also Refs. \cite{lau2018novel,kane1957band,shiozaki2014} therein)].

We investigate the electronic structure of  Na$_2$CdSn using DFT as implemented in the WIEN2k package \cite{wien2k}, a full potential all-electron code employing the linearized augmented plane-wave plus local orbitals method. We use the Perdew-Burke-Ernzerhof (PBE) exchange-correlation functional \cite{Perdew1996PRL}, a Monkhorst-Pack k-grid of 15$\times$15, and self-consistent convergence criteria of 10$^{-6}$ $e$ (elementary charge) for the charge and 10$^{-6}$ Ry for the energy. We consider a core–valence separation energy of $-6$ Ry, atomic spheres with orbital quantum numbers up to 10, and the plane-wave cutoff multiplied by the smallest atomic radii is set to 8. In the absence of spin-orbit coupling (SOC), we optimize the structural parameters and obtain an in-plane lattice parameter $a = 4.978\,$\AA{}  and a thickness (vertical distance between Na atoms) of $3.332\,$\AA{} (consistent with the lattice parameters of the 3D bulk compound 
\cite{peng2018,schuster1980ternare}). In the SM \cite{Na2CdSnsupmat} we present the total energy obtained within DFT as a function of the lattice parameter $a$ and thickness $d$. In the presence of SOC, core electrons are considered fully relativistically, whereas valence electrons are treated in a second variational step \cite{Singh2006}, with the scalar-relativistic wave functions calculated in an energy window of $-10$ to $4$ Ry. We consider a vacuum region of $20\,$\AA.

In Figs. \ref{fig:fig1}-d) and \ref{fig:fig1}-e), we show the electronic band structure and the orbital composition, respectively, obtained via DFT calculations for Na$_2$CdSn in the absence of SOC. The Fermi level ($E_F$) lies at zero energy, and the blue dashed circle indicates the low-energy region we are interested in. The band structure is that of a semimetal with two $\Gamma_6$ bands fourfold degenerate at the $\Gamma$ point (accounting for the twofold spin degeneracy of each band).  The closest remote bands belong to the $\Gamma_1$ and $\Gamma_4$ irreps \cite{Na2CdSnsupmat}. We note that the $\Gamma_6$ bands are dominated by the $p_x$ and $p_y$ orbitals of the Cd and Sn atoms [dark blue in \ref{fig:fig1}-e)], while the $\Gamma_1$ bands are dominated by $s$ orbitals (yellow) with a minor contribution from  $p_x$ and $p_y$ (for more details, see SM \cite{Na2CdSnsupmat}).

In the presence of SOC, Fig. \ref{fig:fig1}-f), the degeneracy at the $\Gamma$ point is lifted, and an energy gap $E_g=234.8\,$meV opens up. The material then becomes a semiconductor with topmost valence bands belonging to the $\Gamma_7$ and $\Gamma_8$ irreps, while the lowest conduction bands belong to $\Gamma_7$ and $\Gamma_9$. The next remote conduction band belongs to the $\Gamma_8$ irrep. All irreps are obtained from the total wave function. 

In Fig. \ref{fig:fig1}-g), we show the expectation value of the $z$ component of spin (color coded) $\langle \hat{S}_z \rangle$ [$\langle \hat{S}_z\rangle>0$ ($\langle \hat{S}_z\rangle<0$) red (blue)].  We note that along the $\Gamma-M$ high-symmetry line ($\Sigma$), all energy bands are doubly degenerate. This high-symmetry line belongs to the point group $C_{2v}$, whose double group has only one irrep -- the 2D $\Sigma_5$ \cite{koster1963book}. The spin-up and down components of the electron wave functions transform according to the two basis functions of $\Sigma_5$. Therefore, all energy bands must be spin degenerate along this direction. Since any linear combination of the spin-degenerate states is also an eigenstate of the Hamiltonian, $\langle \hat{S}_z \rangle$ is not univocally defined along $\Gamma-M$. Hence, we represent the degenerate bands using the black color.

The $\Gamma-K$ ($T$) and $M-K$ ($T'$) lines, on the other hand, belong to the point group $C_s$, which consists of only two elements: the identity and the mirror symmetry $\sigma_h$. Its double group representation has two one-dimensional (1D) irreps, $T_3$ ($T'_3$) and $T_4$ ($T'_4$) for spin-up and -down states, respectively, along the $T$ ($T'$) line. Thus, the energy bands are spin split along these directions. The spin expectation values of all eigenstates out of the $\Gamma-M$ line are given by $\langle \hat{S}_z \rangle \approx \pm \hbar/2$ and $\langle \hat{S}_x \rangle = \langle \hat{S}_y \rangle = 0$, as required by mirror symmetry about the $z$ direction (see Appendix \ref{sec:mirrorsymmetry}) \cite{kurpas2019spin}. We note that our DFT calculated band structure is in great agreement with the previous report of Mao \emph{et al.} \cite{mao2019dual}.

Based on our DFT calculations, in the next section, we focus on the derivation of an effective model to describe the electronic states close to $E_F$ around the $\Gamma$ point.

\section{The effective \kp model}\label{sec:kpmodel}

Following the approach adopted by Liu \emph{et al.} \cite{liu2010model}, we apply the \kp method together with perturbation theory to derive an effective low-energy Hamiltonian for the quasi-2D Na$_2$CdSn (for details, see SM~\cite{Na2CdSnsupmat}).  Our final $4\times4$ model describes the four energy bands close to $E_F$ around the $\Gamma$ point [$\Gamma_8$ and $\Gamma_9$ bands in Fig. \ref{fig:fig1}-f)].

The general \kp Hamiltonian around the $\Gamma$ point is given by \cite{winkler2003book}%
\begin{equation}\label{eq:Hamiltonianterms}
\mathcal{H}(\bm{k})=\epsilon(\bm{k})+\mathcal{H}_0+\mathcal{H}_{\bm{k\cdot p}}+\mathcal{H}_{so}+\mathcal{H}_{\bm{k}so},
\end{equation}
where $\bm{k}$ is the crystal momentum and $\epsilon(\bm{k})=\frac{\hbar^2k^2}{2m_0}$, with $k=|\bm{k}|$, $\hbar$ the reduced Planck constant, and $m_0$ the bare electron mass. The second term in Eq. \eqref{eq:Hamiltonianterms},%
\begin{equation}\label{H0term}
\mathcal{H}_0=\frac{\bm{p}^2}{2m_0}+V(\bm{r}),
\end{equation}%
corresponds to the Hamiltonian at $\bm{k}=0$ in the absence of SOC, with $\bm{p}=-i\hbar\bm{\nabla}$ the linear momentum operator and $V(\bm{r})$ the effective single-particle crystal potential. The \kp and $\bm{k}$-independent SOC terms read, respectively, %
\begin{equation}\label{Hkpterm}
\mathcal{H}_{\bm{k\cdot p}}=\frac{\hbar}{m_0}\bm{k}\cdot\bm{ p},
\end{equation}
and
\begin{equation}\label{HSOterm}
\mathcal{H}_{so}=\frac{\hbar}{4m_0^2c^2}\bm{A}\cdot\hat{\bm{\sigma}},
\end{equation}%
with $\bm{A}=\bm{\nabla}V\times \bm{p}$,  $c$ the speed of light, and $\hat{\bm{\sigma}}\,=(\hat{\sigma}_x,\hat{\sigma}_y,\hat{\sigma}_z)$ the spin operator divided by $\hbar/2$. %
Finally, the $\bm{k}$-dependent SOC term is given by
\begin{equation}\label{Hksoterm}
\mathcal{H}_{\bm{k}so}=\frac{\hbar^2}{4m_0^2c^2}\bm{k}\cdot\hat{\bm{\sigma}}\times\bm{\nabla}V(\bm{r}).
\end{equation}

The perturbative corrections to the $\bm{k}\cdot\bm{p}$ Hamiltonian are calculated by including the remote energy bands via L{\"o}wdin perturbation theory or the folding-down method. We note that, even though the L{\"o}wdin perturbation and folding-down techniques are essentially equivalent, the calculations are performed differently in each case. In the L{\"o}wdin approach, we can include all the infinite remote bands in the perturbative corrections of the \kp Hamiltonian up to second order in the crystal momentum. In the folding down, on the other hand, we can only account for a finite number of remote bands.

Next, we show our effective Hamiltonian derived via L{\"o}wdin perturbation theory. The derivation via folding down is discussed in detail in the SM \cite{Na2CdSnsupmat}.

\subsection{L{\"o}wdin perturbation}\label{subsec:lowdinresults}%

As mentioned above, here we will focus on the topmost valence and lowest conduction bands at the $\Gamma$ point and include the remote bands perturbatively up to second order in $\bm{k}$  (see more details about the L{\"o}wdin perturbation theory in the SM \cite{Na2CdSnsupmat}). From the DFT results, Fig. \ref{fig:fig1}-d), we extract that the bands we are interested in are described by the two basis functions of the $\Gamma_6$ irrep. According to the character table of group $D_{3h}$~\cite{Na2CdSnsupmat}, these eigenstates can be represented by the $|X\rangle$ and $|Y\rangle$ basis functions, which constitute the basis set for the orbital subspace of the truncated Hamiltonian. As usual, we use the eigenstates of the $\hat{S}_z$ operator, $\left|\uparrow\right\rangle$ and $\left|\downarrow\right\rangle$, as the basis for the spin subspace. Finally, the four-dimensional spinful basis set is given by the direct product between the basis functions of the spin subspace and those of the orbital subspace, e.g., $\left|X\!\uparrow\right\rangle= \ket{\uparrow} \otimes \ket{X} $. 

The expressions for the first-order terms are obtained by projecting the total Hamiltonian [Eq.~(\ref{eq:Hamiltonianterms})] onto a truncated basis consisting of eigenstates of $\mathcal{H}_0$ [Eq. (\ref{H0term})]. The matrix elements of the second-order contributions, on the other hand, are calculated using standard perturbation theory formulas (Eq. (S30) in the SM~\cite{Na2CdSnsupmat}). In practice, the matrix representation of the second-order contributions is calculated separately and added to the first-order matrix Hamiltonian:

\begin{equation}\label{eq:HamiltonianLowdin}
\begin{split}
    H=&\epsilon^{(1)} (\bm{k})+H^{(1)}_0+H^{(1)}_{\bm{k}\cdot\bm{p}}+H^{(1)}_{so}+H^{(1)}_{\bm{k}so} \\ 
    &+ H_{\bm{k}\cdot\bm{p}}^{(2)} + H_{\bm{k}\cdot\bm{p},so}^{(2)} +  H_{so,\bm{k}\cdot\bm{p}}^{(2)} \\
    &+ H_{\bm{k}\cdot\bm{p},\bm{k}so}^{(2)} + H_{\bm{k}so,\bm{k}\cdot\bm{p}}^{(2)} \\
    & + H_{so}^{(2)} +  H_{so,\bm{k}so}^{(2)}+H_{\bm{k}so,so}^{(2)} + H_{\bm{k}so}^{(2)} .
\end{split}
\end{equation}

\begin{table}[h]
	\caption{First-order coupling parameters of the \kp Hamiltonian in Eq. ~\eqref{eq:HamiltonianLowdin}. Left and right columns show the parameters and their expressions, respectively. All parameters are real. $\varepsilon_{ij3}$ is the Levi-Civita (permutation) symbol, and repeated indices are summed over, according to the Einstein convention. For more details on the group theory calculations, see Sec. S-V of SM \cite{Na2CdSnsupmat}.}
	\centering
	\begin{tabularx}{\linewidth}{l@{\extracolsep{\fill}}c}
		\hline
		\hline
		Parameter & Expression \\
		\hline
		$\epsilon_0$ & $\left\langle X \middle | \mathcal{H}_0 \middle | X \right\rangle$ \\
		$\alpha$ & $-\hbar^2\left\langle X\middle | \varepsilon_{ij3}\partial_iV\partial_j\middle| Y\right\rangle/\left(4m_0^2c^2\right)$ \\
		$\zeta$ & $\hbar^2\left\langle X\middle| \partial_xV\middle| X\right\rangle/\left(4m_0^2c^2\right)$\\
		\hline
		\hline
	\end{tabularx}
	\label{tab:parametersfolding}
\end{table}

Projecting each term of Eq. (\ref{eq:Hamiltonianterms}) onto the basis $\{|X\uparrow\rangle$, $|Y\uparrow\rangle$, $|X\downarrow\rangle$, $|Y\downarrow\rangle\}$  and employing group theory analysis to find selection rules for the matrix elements (see Sec. S-V in the SM \cite{Na2CdSnsupmat}) yields%
\begin{gather}
    \epsilon^{(1)}(\bm{k})=\epsilon(\bm{k})\sigma_0\otimes\tau_0,\\
     \label{eqH0} H^{(1)}_0=\epsilon_0\sigma_0\otimes\tau_0, \quad H^{(1)}_{\bm{k}\cdot \bm{p}}=0,
\end{gather}%
\begin{gather}
    H^{(1)}_{so}=-\alpha\sigma_3\otimes\tau_2,\\
    H^{(1)}_{\bm{k}so}=\zeta\sigma_3\otimes \left(k_x\tau_1+k_y\tau_3\right), \label{Hkso_1}
\end{gather}
where $\tau_0$ $(\sigma_0)$ and $\tau_i$ $(\sigma_i)$, $ i=1,2,3$, are the $2\times2$ identity and Pauli matrices in the orbital (spin) subspace, respectively. The expressions for the parameters of the first-order terms are provided in Table \ref{tab:parametersfolding}.

For the second-order contributions, we have
\begin{widetext}%
\begin{gather} \label{eq:2ndorder1}
H^{(2)}_{\bm{k\cdot p}}= \frac{1}{2}\sigma_0\otimes\left[\left(A_1+A_2\right)k^2\tau_0%
+2\left(A_1-A_2\right)k_xk_y\tau_1%
+\left(A_1-A_2\right)\left(k_x^2-k_y^2\right)\tau_3\right],\\
H^{(2)}_{\bm{k\cdot p},so}+H^{(2)}_{so,\bm{k\cdot p}}=D_1\sigma_3\otimes\left(k_x\tau_1+k_y\tau_3\right),\quad\quad\quad\quad\quad
H^{(2)}_{\bm{k\cdot p},\bm{k}so}+H^{(2)}_{\bm{k}so,\bm{k\cdot p}}=%
Bk^2\sigma_3\otimes\tau_2,\\
H^{(2)}_{so}=C_1\sigma_0\otimes\tau_0-C_2\sigma_3\otimes\tau_2,\quad\quad
H^{(2)}_{so,\bm{k}so}+H^{(2)}_{\bm{k}so,so}=E_1\sigma_3\otimes\left(k_x\tau_1+k_y\tau_3  \right),\\
H^{(2)}_{\bm{k}so}= \frac{1}{2}\sigma_0\otimes \left[\left(F_1+F_2\right)k^2\tau_0
+2\left(F_1-F_2\right)k_xk_y\tau_1+\left(F_1-F_2\right)\left(k_x^2-k_y^2\right)\tau_3\right],
 \label{eq:2ndorder2}
\end{gather}
\end{widetext}
with second-order parameters $A_1$, $A_2$, $D_1$, $B$, $C_1$, $C_2$, $E_{1}$, $F_1$ and $F_2$ given in Appendix \ref{sec:expressionsparametersLowdin}.
Note that, here, we have considered the influence of all the remote bands up to second order in $\bm{k}$. Equations~\eqref{eqH0}--\eqref{eq:2ndorder2} are in agreement with our derivation using the theory of invariants (see Appendix~\ref{sec:theoryofinvariants}).

The eigenvalues of the Hamiltonian in Eq. (\ref{eq:HamiltonianLowdin}) at $\bm{k}=\bm{0}$ read $\tilde{E}_1=\tilde{E}_3= C+M$ and $\tilde{E}_2= \tilde{E}_4= C-M$, with $C=\epsilon_0+C_1$ and $M=\alpha+C_2$. The corresponding eigenvectors are given by $|u_{1(2),\bm{0}}\uparrow\rangle$ and $|u_{3(4),\bm{0}}\downarrow\rangle$ in Table \ref{tab:evecsfoldingdown}. Next, we follow a usual procedure in the \kp literature \cite{winkler2003book,Calsaverini2008intersubband, fu2020spinorbit} and perform a change of basis from $\{|X\uparrow\rangle$, $|Y\uparrow\rangle$, $|X\downarrow\rangle$, $|Y\downarrow\rangle\}$ to $\{|u_{1,\bm{0}}\uparrow\rangle$, $|u_{2,\bm{0}}\uparrow\rangle$, $|u_{3,\bm{0}}\downarrow\rangle$, $|u_{4,\bm{0}}\downarrow\rangle\}$, rewriting the full $\bm{k}$-dependent Hamiltonian in a simpler form (diagonal at $\bm{k}=0$):
\begin{widetext}%
\begin{equation}\label{blockdiagonalhamiltonian}
H(\bm{k})=%
\begin{pmatrix}
C+M-(D+B)k^2 & -iAk_-+Gk_+^2   & 0 & 0 \\
  iAk_++Gk_-^2 & C-M-(D-B)k^2 & 0 & 0\\
0 & 0 & C+M-(D+B)k^2 & -iAk_++Gk_-^2 \\
0 & 0 & iAk_-+Gk_+^2 & C-M-(D-B)k^2
\end{pmatrix},%
\end{equation}%
\end{widetext}%
where
\begin{align}
    A=&D_1+E_1+\zeta,\label{eq:paramA}\\
    D=&-\frac{1}{2}\left[\left(A_2+A_1\right)+\left(F_2+F_1\right)+\frac{\hbar^2}{m_0}\right]\label{eq:paramD},
\end{align}%
and%
\begin{equation}
    G=\frac{1}{2}\left[\left(A_2-A_1\right)+\left(F_2-F_1\right) \right].\label{eq:paramG}
\end{equation}

\begin{table}[h]
	\caption{Truncated set of zone-center wave functions used as the basis of the Hamiltonian in Eq.  (\ref{blockdiagonalhamiltonian}).}
	\centering
	\begin{tabularx}{\linewidth}{l@{\extracolsep{\fill}}c}
		\hline
		\hline
		Basis function & Expression\\
		\hline
		$|u_{1,\bm{0}}\uparrow\rangle$ &  $ \frac{i}{\sqrt{2}}(|X\uparrow\rangle- i|Y\uparrow\rangle)$\\
		$|u_{2,\bm{0}}\uparrow\rangle$ &  $- \frac{i}{\sqrt{2}}(|X\uparrow\rangle+ i|Y\uparrow\rangle)$\\
		$|u_{3,\bm{0}}\downarrow\rangle$ & $-\frac{i}{\sqrt{2}}(|X\downarrow\rangle+ i|Y\downarrow\rangle)$\\
		$|u_{4,\bm{0}}\downarrow\rangle$ & $\frac{i}{\sqrt{2}}(|X\downarrow\rangle- i|Y\downarrow\rangle)$\\
		\hline
		\hline
	\end{tabularx}
	\label{tab:evecsfoldingdown}
\end{table}

The matrix Hamiltonian in Eq. (\ref{blockdiagonalhamiltonian}) is block diagonal, with each block corresponding to a different spin. This is in total agreement with the derivation via folding down (Sec. S-V in the SM \cite{Na2CdSnsupmat}).
Since there is no coupling between these two blocks, the $z$ component of spin is a good quantum number. As we shall see in Sec. \ref{sec:fitting}, the parameter $B$, arising from the  $H_{\bm{k\cdot p},\bm{k}so}^{(2)}$ and $H_{\bm{k}so,\bm{k\cdot p}}^{(2)}$ terms often neglected for other materials in the literature, is very important for the topology of our system.

Our \kp Hamiltonian contains off-diagonal matrix elements $Gk_\pm^2$ (absent in the model of Mao \emph{et al.} \cite{mao2019dual}), which are captured by keeping all contributions up to second order in $\bm{k}$ in the Löwdin perturbation theory, folding-down technique, or theory of invariants (Appendix \ref{sec:theoryofinvariants}). In addition, the Hamiltonian proposed by Mao \emph{et al.} \cite{mao2019dual} has terms linear in $k_\pm$ in the secondary diagonal, coupling opposite spins. In our model, any term connecting the two spin subspaces is forbidden by symmetry \footnote{By applying the theory of invariants using an extended basis set, we have verified that the next remote band, $\Gamma_4$, couples to the bands of interest, $\Gamma_6$, through zeroth- and first-order in $\bm{k}$ spin-mixing terms. However, after eliminating the remote bands via Löwdin perturbation theory, the final $4\times 4$ Hamiltonian remains equivalent to Eq. (\ref{blockdiagonalhamiltonian}), with the new terms merely renormalising the model parameters. This result is in line with our model derivation in Appendix \ref{sec:theoryofinvariants}, where we have obtained all terms allowed by symmetry
through the theory of invariants.}. This result was obtained via all the three perturbative approaches considered in this paper.

In the next section, we analyze the symmetries of our effective \kp Hamiltonian.

\subsection{Symmetry analysis}\label{symmanalysis}

The TRS operator $\mathcal{T}=-i\hat{\sigma}_{y}K$, with $K$ the complex conjugate operator, projected onto the basis $\left\{|u_{1,\bm{0}}\uparrow\rangle,|u_{2,\bm{0}}\uparrow\rangle,|u_{3,\bm{0}}\downarrow\rangle,|u_{4,\bm{0}}\downarrow\rangle\right\}$ (Table \ref{tab:evecsfoldingdown}) is given by $T=-i\sigma_{2}\otimes\tau_{0}K$. We have verified that our model Hamiltonian in Eq. (\ref{blockdiagonalhamiltonian}) respects the condition of invariance under TRS by satisfying $T H\left(\bm{k}\right) T^{-1}=H\left(-\bm{k}\right)$.

The mirror symmetry operator for spinful electrons satisfies \cite{teo2008surface,Ando2015}%
\begin{equation}
\hat{\sigma}_h^{2}=-1.
\end{equation}
Hence the eigenvalues of $\hat{\sigma}_h$ are $i$ and $-i$. For a 2D crystal invariant under mirror symmetry about the $xy$-plane, $[\mathcal{H},\hat{\sigma}_h]=0$, the eigenstates of the Bloch Hamiltonian can be chosen to also be eigenstates of $\hat{\sigma}_h$ for all crystal momenta $\bm{k}$. This yields two classes of Bloch eigenstates with mirror eigenvalues $\eta=\pm i$,  important for the calculation of the mirror Chern number (Sec. \ref{sec:topologyanalysis}).

Using the basis of our \kp model (Table \ref{tab:evecsfoldingdown}), the matrix representation of the mirror symmetry operator reads
\begin{equation}\label{eq:mirrorsymmetryoperator}
\sigma_h=-i\sigma_{3}\otimes\tau_{0}=\begin{pmatrix}-i & 0 & 0 & 0\\
0 & -i & 0 & 0\\
0 & 0 & i & 0\\
0 & 0 & 0 & i
\end{pmatrix}.
\end{equation}%
The model Hamiltonian in Eq.~\eqref{blockdiagonalhamiltonian} satisfies~\cite{Na2CdSnsupmat}%
\begin{equation}\label{eq:mirrorinv}
\sigma_hH\left(\bm{k}\right)\sigma_h^{-1}=H\left(u_{\sigma_h}\bm{k}\right),
\end{equation}
where $u_{\sigma_h}$ is the matrix representation of the mirror operator acting on the coordinate space. Therefore the spin texture of the simultaneous eigenstates of $H(\bm{k})$ and $\sigma_h$ must be polarized along the $z$ direction in the whole 2D BZ, except at degeneracies [see Fig. \ref{fig:fig1}-g) and Appendix \ref{sec:mirrorsymmetry}] \cite{kurpas2019spin}. An equation analogous to Eq. (\ref{eq:mirrorinv}) holds for the remaining symmetry operations of the point group $D_{3h}$ (see SM~\cite{Na2CdSnsupmat}).

Having the final effective Hamiltonian [Eq.~\eqref{blockdiagonalhamiltonian}] at hand, we now obtain the corresponding energy bands and perform a fitting procedure using the DFT results [see Fig. \ref{fig:fig1}-f)].

\subsection{Fitting}\label{sec:fitting}

The four energy bands obtained by diagonalizing  the Hamiltonian in Eq. (\ref{blockdiagonalhamiltonian}) are given by%
\begin{align}
E_{1(3),\bm{k}}=&C-Dk^2+\sqrt{M^2+\Delta_\mp(\bm{k})},\label{eq:bulkband1}\\
E_{2(4),\bm{k}}=&C-Dk^2-\sqrt{M^2+\Delta_\pm(\bm{k})},\label{eq:bulkband2}
\end{align}
where $E_{1,\bm{k}}$ and $E_{2,\bm{k}}$ correspond to the spin-up sector, $E_{3,\bm{k}}$ and $E_{4,\bm{k}}$ to the spin-down sector, and%
\begin{equation}
\begin{split}\label{eq:deltapm}
\Delta_\pm&(\bm{k})=\pm 2AG\left(-3k_x^2+k_y^2\right)k_y\\
&+(A^2-2MB)k^2+\left(G^2+B^2\right)k^4.
\end{split}
\end{equation}%
By fitting the energy bands [Eqs. (\ref{eq:bulkband1}) and (\ref{eq:bulkband2})] to the band structure obtained via DFT calculations, we obtain numerical values for the model parameters, shown in Table \ref{tab:parametercontributions}.

A range of $\sim 5.5\%$ of the $\Gamma-M$ line was used in the fitting performed with the LMFIT package for Python \cite{newville_2015_11813}. The LMFIT package provides several minimization methods that usually yield different parameter sets. We followed the iterative procedure used in Ref. \cite{faria2016realistic}, where the best parameter set is chosen after an initial fit and used as an input for a new fit using all minimization methods. The best parameter set is chosen again from the new output and this process is repeated until the parameters that reproduce best the DFT data are obtained.

\begin{figure*}[t]
    \centering
    \includegraphics[width=\linewidth]{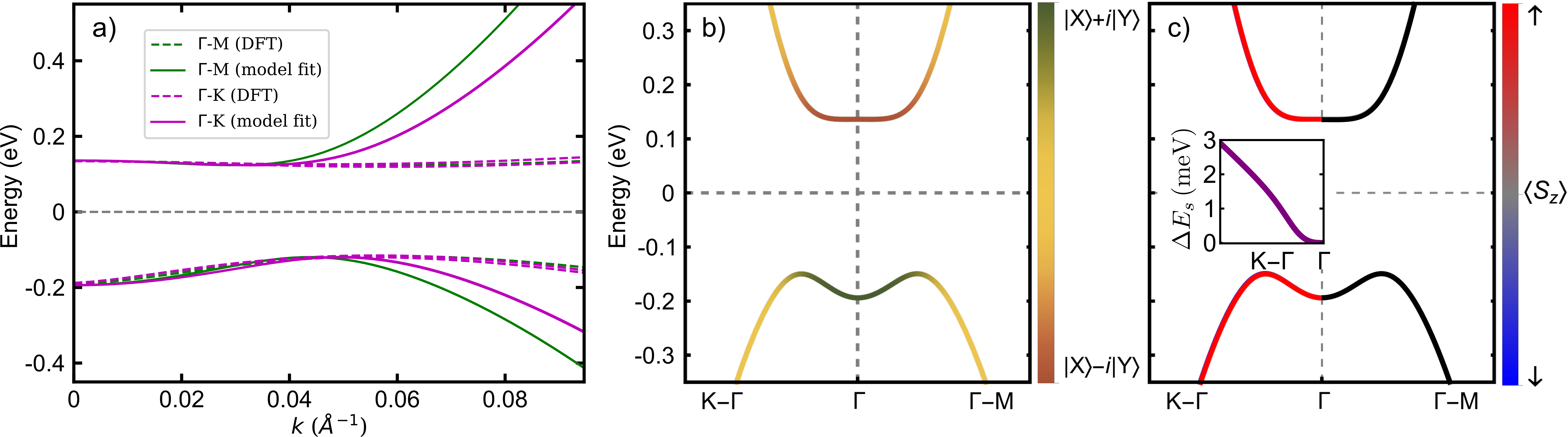}
    \caption{(Color online) Effective energy bands for Na$_2$CdSn. a) Fitting of the $4\times 4$ model Hamiltonian derived via \kp theory to the density functional theory (DFT) energy bands. A range of $\sim 5.5\%$ of the $\Gamma-M$ line was used in the fitting, whose numerical values of the parameters are given in Table \ref{tab:parametercontributions}. Solid (dashed) curves represent the energy bands from the effective model (DFT calculations), and green (magenta) represents the $\Gamma-M$ ($\Gamma-K$) high-symmetry line. (b) Spin-up bands with orbital (pseudospin) composition. The color bar shows the contribution from the $|X\rangle+i|Y\rangle$ (green) and $|X\rangle-i|Y\rangle$ (brownish red) states. (c) The four energy bands of the model (both spins) with spin expectation value. The spin-up bands are colored red, while the spin-down bands are blue. The bands are spin degenerate along the $\Gamma-M$ high-symmetry line. In this case, the spin expectation value is not univocally defined. For this reason, we color the bands black in this direction. Along $\Gamma-K$, there is a spin splitting (not distinguishable in the main graph) due to spin-orbit coupling (SOC), $\Delta E_s$, which is presented in the inset as a function of the crystal momentum.}
    \label{fig:fig2}
\end{figure*}

In Table \ref{tab:parametercontributions}, we also show the corresponding Hamiltonian terms that contribute to each of the model parameters (see Appendix \ref{sec:expressionsparametersLowdin} for details). For $G=0$, the  Hamiltonian in Eq. (\ref{blockdiagonalhamiltonian}) resembles the well-known Bernevig-Hughes-Zhang (BHZ) model \cite{bernevig2006quantum}. In both models, the energy bands are inverted if $MB>0$, and therefore, the parameter $B$ is important for the topological analysis of the band structures. We notice that the contributions to the parameter $B$, accounting for all the remote bands perturbatively, come from $H_{\bm{k\cdot p},\bm{k}so}^{(2)}+H_{\bm{k}so,\bm{k\cdot p}}^{(2)}$, i.e., the second-order contribution via $\mathcal{H}_{\bm{k\cdot p}}$ and $\mathcal{H}_{\bm{k}so}$. This suggests that the second-order correction coming from the $\bm{k}$-dependent SOC should not be neglected for Na$_2$CdSn.

\begin{table}[h]
    \caption{Parameters of the model Hamiltonian derived via Löwdin perturbation theory. The second column shows the numerical values obtained from the fitting to the DFT energy bands, while the third column shows which Hamiltonian terms contribute to each parameter.}
    \centering
    \begin{tabularx}{\linewidth}{l@{\extracolsep{\fill}}r@{\extracolsep{\fill}}c}
    \hline\hline
        &  Fitting value & Contributions \\
       \hline %
       $C$ & $-0.029\,$eV  & $H_0;H_{so}^{(2)}$\\
       $M$ & $0.165\,$eV & $H_{so};H_{so}^{(2)}$\\
       $A$ & $0.0199\,$eV$\cdot$\AA & $H_{\bm{k}so};H_{\bm{k\cdot p},so}^{(2)};H_{so,\bm{k\cdot p}}^{(2)};H_{so,\bm{k}so}^{(2)};H_{\bm{k}so,so}^{(2)}$\\
       $G$ & $-116\,$eV$\cdot$\AA$^2$ & $H_{\bm{k\cdot p}}^{(2)};H_{\bm{k}so}^{(2)}$\\
       $D$ & $-44.33\,$eV$\cdot$\AA$^2$ & $H_{\bm{k\cdot p}}^{(2)};H_{\bm{k}so}^{(2)}$\\
       $B$ & $47.87 \,$eV$\cdot$\AA$^2$ & $H_{\bm{k\cdot p},\bm{k}so}^{(2)}$;$H_{\bm{k}so,\bm{k\cdot p}}^{(2)}$\\
       \hline
       \hline
    \end{tabularx}
    \label{tab:parametercontributions}
\end{table}

In Fig. \ref{fig:fig2}-a), we show the energy bands using the parameters obtained from the fitting (solid curves) and from DFT calculations (dashed). The color code represents the direction in the BZ, with green and magenta for the high-symmetry lines $\Gamma-M$ and $\Gamma-K$, respectively. As expected, our \kp model reproduces well the  DFT data around the $\Gamma$ point but deviates from the numerical data for larger values of the crystal momentum. In addition, note that the energy bands of the analytical model deviate from the DFT bands before they reach the minimum and maximum of the conduction and valence bands, respectively. As a result, the energy gap for the fitted parameters $E_g=285$~meV is $\sim 21\%$ larger than the gap obtained via DFT calculations. This, however, does not affect the topological properties of the material (see Sec. \ref{sec:topologyanalysis}).

In Fig. \ref{fig:fig2}-b), we show the spin-up conduction and valence bands [$E_{1,\bm{k}}$ and $E_{2,\bm{k}}$ in Eqs. (\ref{eq:bulkband1}) and (\ref{eq:bulkband2}), respectively] for the fitted parameters with the orbital composition (color code). Note that the conduction band is dominated by the $|X\rangle-i|Y\rangle$ state (brownish red) at the $\Gamma$ point, while the valence band is dominated by $|X\rangle+i|Y\rangle$ (green). For larger values of the crystal momentum, both states become relevant to the two energy bands.

In Fig. \ref{fig:fig2}-c), we show the four energy bands for the fitted parameters, colored according to their spin texture. The bands are spin degenerate along the $\Gamma-M$ direction, in agreement with the DFT results (see Fig. \ref{fig:fig1}). The spin splitting along the $\Gamma-K$ line, $\Delta E_s\equiv E_{3,\bm{k}}-E_{1,\bm{k}}=E_{4,\bm{k}}-E_{2,\bm{k}}$, is very small, reaching a maximum of $3\,$meV in the range shown. In the inset we display $\Delta E_s$ as a function of $\bm{k}$. Note that this spin splitting is not accompanied by a mixing of the spin components since there is no coupling between spin-up and spin-down blocks in the effective Hamiltonian in Eq. (\ref{blockdiagonalhamiltonian}). According to Eq. (\ref{eq:deltapm}),  such a spin splitting requires that both parameters $A$ and $G$ be finite.

In the next section, we perform a topological analysis of our effective model and predict that the Na$_2$CdSn is a DTI characterized by nontrivial topological invariants.

\section{Topology analysis}\label{sec:topologyanalysis}

Our model Hamiltonian in Eq. (\ref{blockdiagonalhamiltonian}) can be rewritten in the compact form%
\begin{equation}\label{eq:compactform}%
H(\bm{k})=%
\begin{pmatrix}%
h(\bm{k}) & 0 \\
0 & h^*(-\bm{k})%
\end{pmatrix},%
\end{equation}%
with $h(\bm{k})=d_0(\bm{k})\bar{\tau}_0+\bm{d}(\bm{k})\cdot \bm{\bar{\tau}}$. The matrices  $\bar{\tau}_0$ and $\bm{\bar{\tau}}=\left(\bar{\tau}_1,\bar{\tau}_2,\bar{\tau}_3\right)$ are the identity and Pauli matrices in the basis $\{i\left(|X\rangle-i|Y\rangle\right)/2,-i\left(|X\rangle+i|Y\rangle\right)/2\}$ of the orbital subspace (pseudospin degree of freedom), respectively. The function $d_0(\bm{k})$ and the components of the pseudospin vector  $\bm{d}(\bm{k})=\left[d_1(\bm{k}),d_2(\bm{k}),d_3(\bm{k})\right]$ are given by $d_0(\bm{k})=C-Dk^2$, $d_1(\bm{k})=-Ak_y+G(k_x^2-k_y^2)$, $d_2(\bm{k})=Ak_x-2Gk_xk_y$, and $d_3(\bm{k})=M-Bk^2$. The spin-up and down blocks [$h(\bm{k})$ and $h^*(-\bm{k})$, respectively] are connected by TRS.

To investigate in more detail the topology of Na$_2$CdSn, we perform a square lattice regularization on Eq. (\ref{eq:compactform}), which consists of substituting $k_{i} \rightarrow \sin\left(ak_{i}\right)/a$ and $k_{i}^{2} \rightarrow 2\left[1-\cos\left(ak_{i}\right)\right]/a^{2}$ ($i=x,y$) \cite{bernevig2013book}. This substitution is valid for $|k_i|\ll 1/a$ and, therefore, expected to be valid in our region of interest.

\begin{figure*}
    \centering
    \includegraphics[width=\linewidth]{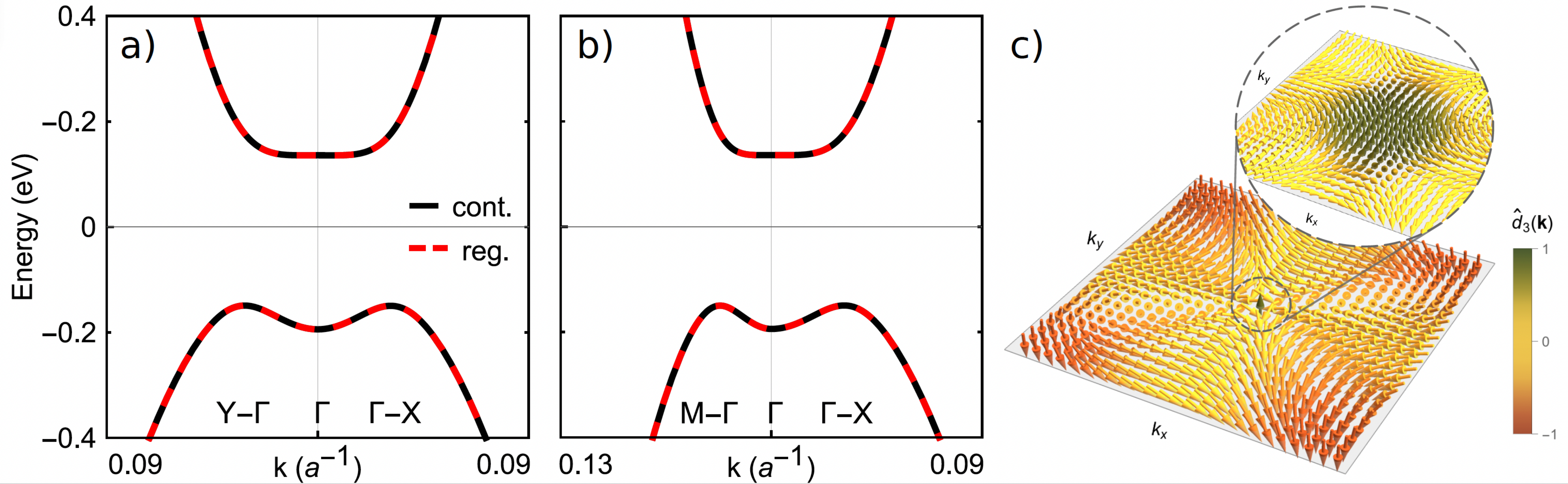}
    \caption{(Color online) Comparison between spin-up energy bands for the continuous (solid black curves) and regularized (dashed red curves) models along $3\%$ of the a) $Y-\Gamma-X$ and b) $M-\Gamma-X$ high-symmetry paths of the square-lattice Brillouin zone (BZ), corresponding to the region of interest. The crystal momentum intervals along each direction are given by $0\leq|k|<0.09/a$  along either $\Gamma-X$ or $\Gamma-Y$, and $0\leq|k|<0.03 \sqrt{2}\pi/a$ along the $\Gamma-M$ direction. c) Pseudospin skyrmion configuration of the regularized model over the first BZ, i.e., $0<|k_{x(y)}|<\pi/a$. Each arrow is determined by the components of the pseudospin unit vector at the corresponding $\bm{k}$ point, $\hat{\bm{d}}(\bm{k})$, with the color code given by the $\hat{d}_3(\bm{k})$ component. The inset shows a zoom into the region of interest, where we can see in more detail the behavior of the vector field around the $\Gamma$ point. The ranges for the components of the crystal momentum in the inset are given by $0<|k_{x(y)}|<0.02\pi/a$.}
    \label{fig:regplots}
\end{figure*}

Using the parameters in Table \ref{tab:parametercontributions}, we show in Fig. \ref{fig:regplots}-a) the spin-up energy bands along the high-symmetry path $Y-\Gamma-X$ for both continuous and regularized models. Here, $\Gamma=(0,0)$, $X=(\pi/a,0)$, $Y=(0,\pi/a)$ and $M=(\pi/a,\pi/a)$ are high-symmetry points of the first BZ for the square lattice. The energy bands are plotted only within the region of interest, $\sim$3\% of the high-symmetry lines, i.e., $0\leq|k|\lesssim 0.09\,a^{-1}$.  The black solid and red dashed curves represent the energy bands of the continuous and regularized models, respectively. In Fig. \ref{fig:regplots}-b) we show the spin-up energy bands along $3\%$ of the $M-\Gamma-X$ high-symmetry path, which for the $\Gamma-M$ direction corresponds to $0\leq|k|\lesssim 0.13\,a^{-1}$.

 The excellent agreement between the continuous and regularized models around the  $\Gamma$ point is clear from Figs. \ref{fig:regplots}-a) and \ref{fig:regplots}-b). In addition, we have confirmed (by plotting the energy bands along the full path, not shown) that the regularization procedure preserves the overall gap, which is important to calculate an integer-valued topological invariant \cite{bernevig2013book}.

The pseudospin unit vector for the regularized model  $\hat{\bm{d}}(\bm{k})=\bm{d}(\bm{k})/|\bm{d}(\bm{k})|$ defines a mapping from the compact BZ to the unit Bloch sphere. The number of times this mapping wraps the Bloch sphere defines an integer-quantized topological invariant known as the winding number \cite{qi2006topological,Sticlet2012,bernevig2013book}. In Fig. \ref{fig:regplots}-c), we show the skyrmion configuration obtained for Na$_2$CdSn. Each vector is defined by the three components of $\hat{\bm{d}}(\bm{k})$ at the corresponding $\bm{k}$ point of the BZ, and the color code denotes the $\hat{d}_3(\bm{k})$ component. We can see that, at the $\Gamma$ point, the vector field points in the positive $z$ direction, while at the corners of the BZ, the vectors point in the negative $z$ direction. Additionally, the skyrmion configuration in Fig. \ref{fig:regplots}-c) resembles that of a second-order antiskyrmion, which alludes to a topological charge of $-2$ \cite{gobel2021beyond,nagaosa2013topological,camosi2018micromagnetics,muller2018thesis}.

To unequivocally identify the topological phase of the regularized model, we calculate the Chern number for the valence bands of each spin sector separately. The first step is to define the Berry connection $\bm{A}_v^{\uparrow(\downarrow)}(\bm{k})=-i\langle u_v^{\uparrow(\downarrow)}(\bm{k})|\bm{\nabla}_{\bm{k}}|u_v^{\uparrow(\downarrow)}(\bm{k})\rangle$ and Berry curvature $F^{\uparrow (\downarrow)}_v(\bm{k})=\hat{z}\cdot\left[\bm{\nabla}_{\bm{k}}\times \bm{A}_v^{\uparrow(\downarrow)}(\bm{k}) \right]$ \cite{bernevig2013book}, where $|u_v^{\uparrow (\downarrow)}(\bm{k})\rangle$ denotes the valence band of the corresponding spin sector. Now we define the Chern number for each spin sector using the well-known TKNN formula \cite{thouless1982quanzited}, which consists of integrating the Berry curvature over the whole BZ \cite{bernevig2013book}:%
\begin{equation}\label{eq:Chernnumberspinsectors}
C^{\uparrow (\downarrow)}=\frac{1}{2\pi}\int_{BZ}d^2kF^{\uparrow(\downarrow)}_v(\bm{k}).
\end{equation}%
Due to TRS, the total Chern number must be zero, i.e., $C=\left(C^\uparrow+C^\downarrow\right)=0$. By calculating the Chern numbers numerically, we obtain $C^\uparrow=-2$ and $C^\downarrow=2$ for our system, consistent with the second-order antiskyrmion configuration in Fig. \ref{fig:regplots}-c). This  is  related to the presence of the $Gk_\pm^2$ terms in the off-diagonal matrix elements of the \kp Hamiltonian in Eq. (\ref{blockdiagonalhamiltonian}). For $G = 0$ (with the remaining parameters unchanged), the Chern numbers are given by $C^\uparrow=1$ and $C^\downarrow=-1$.

An important topological invariant used to characterize 2D TIs is the so-called spin Chern number \cite{sheng2005nondissipative,sheng2006quantum,fukui2007topological,prodan2009robustness,li2010chern,yang2011time,li2013spin,lv2021measurement}, defined as%
\begin{equation}
    C_s=\frac{1}{2}\left(C^\uparrow-C^\downarrow\right).
\end{equation}
For our system, $C_s=-2$, which characterizes the presence of a TI phase. Note that, at zero temperature, the spin Hall conductivity within the bulk band gap is related to the spin Chern number via $\sigma_s = C_s (e/2\pi)$ \cite{kane2005quantum}. Therefore, we predict that $\sigma_s = -2 e/2\pi$ for Na$_2$CdSn.

We now calculate the mirror Chern number \cite{teo2008surface}, a topological invariant often used to classify TCIs protected by mirror symmetry \cite{liu2014spin, Ando2015,liu2015crystal,hsu2016two}. Since the mirror operator for fermions satisfies $\hat{\sigma}_h^2=-1$, its eigenvalues are given by $\pm i$ (see Appendix \ref{sec:mirrorsymmetry}). Due to the simple diagonal matrix representation of the mirror operator in our model [Eq. (\ref{eq:mirrorsymmetryoperator})], we can readily label spin-up (spin-down) energy bands with the mirror eigenvalue $-i$ ($+i$). In analogy to the spin Chern number, the mirror Chern number can be defined as \cite{teo2008surface}%
\begin{equation}
    C_m=\frac{1}{2}\left(C^{-i}-C^{+i}\right),
\end{equation}%
where $C^{\pm i}$ is obtained by substituting $\uparrow (\downarrow)\rightarrow +i (-i)$ in Eq. (\ref{eq:Chernnumberspinsectors}). For our system, $C_m=-2$.

As shown above, Na$_2$CdSn hosts the TI and TCI phases simultaneously; thus, it can be classified as a DTI protected by TRS and mirror symmetry. Note that both topological invariants (spin and mirror Chern numbers) are equal to $-2$ (in sharp contrast with the spin and mirror Chern numbers $-1$ reported in Ref. \cite{mao2019dual}). According to the bulk-boundary correspondence \cite{qi2006general,mong2011edge}, the Na$_2$CdSn triatomic layer in a finite geometry must have two pairs of counterpropagating helical edge states at each of its boundaries \cite{qi2006topological}. We explore this feature in the next section.

Unlike well-known conventional TIs \cite{bernevig2006quantum,Zhang2009}, the band inversion in Na$_2$CdSn does not follow the closure of a trivial bulk band gap. Here, the topological gap arises from the SOC-induced lifting of a zone-centered nodal point [Figs. \ref{fig:fig1}-d) and \ref{fig:fig1}-e)]. The nodal point degeneracy is enforced by the 2D irrep $\Gamma_6$ of the low-energy conduction and valence bands. Once SOC is considered, the degeneracy is immediately lifted [Figs. \ref{fig:fig1}-f) and \ref{fig:fig1}-g)], resulting in a topologically nontrivial band gap characterized by the spin and mirror Chern numbers -2.

We note that the topology of Na$_2$CdSn is fundamentally distinct from reported systems with similar crystal structures, such as Na$_2$MnPb and $\alpha$-Bi. The former has been predicted to be a ferromagnetic DTI with $C_s=1$  and broken TRS \cite{wang2020ferromagnetic}, while the latter is predicted to have $C_s=2$ but no crystal-symmetry-related topological invariant \cite{bai2022doubled}. Additionally, it is noteworthy that Ref. \cite{qi2006topological} reported a nontrivial spin Chern number $C_s=2$ in a generic model for 2D mirror-symmetric paramagnetic semiconductors. However, the mirror Chern number, proposed in Ref. \cite{teo2008surface}, was not investigated in their study.

\section{Topological edge states}\label{sec:edgestates}

Let us now investigate the existence of topological edge states predicted by the nontrivial topological invariants calculated in Sec. \ref{sec:topologyanalysis}. In Sec.~\ref{sec:semiinfiniteplane}, we analytically obtain the edge states for a semi-infinite plane geometry. In Sec.~\ref{sec:ribbongeometry}, we perform a numerical study of the topological edge states in a ribbon geometry.

\subsection{Semi-infinite plane}\label{sec:semiinfiniteplane}

We first consider a semi-infinite plane geometry consisting of vacuum and the DTI material occupying the $x<0$ and $x>0$ half-planes, respectively. Such geometry can be modeled by a piecewise potential given by%
\begin{equation}
V(x)=%
\begin{cases}
0 & x \geq 0,\\
+\infty & x < 0.
\end{cases}
\end{equation}%
Note that the system is no longer periodic along the $x$ direction, so $k_x$ is not a good quantum number.  On the other hand, translation symmetry is preserved along the $y$ direction, so $k_y$ is still a good quantum number. We then employ the envelope function approach and perform the substitution $k_x\rightarrow -i\partial_x$ in the \kp model Hamiltonian~\cite{bastard1990book,michetti2012helical}. 

In this geometry, any edge-localized eigenstate $\Psi_{k_y}(x,y)$ must satisfy the following boundary conditions:
\begin{align}
    \Psi_{k_y}(x=0,y)=\,0 \quad &\forall\, y\in \mathbb{R}\label{bondcond1},\\
    \lim_{x\rightarrow +\infty}\Psi_{k_y}(x,y)=\,0 \quad &\forall\, y\in \mathbb{R},\label{bondcond2}\\
    \Psi_{k_y}(x,y+L_y)=\,\Psi_{k_y}(x,y) \quad &\forall\, (x,y)\in \mathbb{R}^2.\label{bondcond3}
\end{align}%
Equation (\ref{bondcond3}) is the Born-von Karman boundary condition, with $L_y$ the length of the system in the $y$ direction.

Consider the Schr{\"o}dinger equation for the spin-up block $h(k_x\rightarrow-i\partial_x,k_y)\psi(x,y)=E\psi(x,y)$, where $h(\bm{k})$ is defined by Eq. (\ref{eq:compactform}) and $\psi(x,y)$ is a particular two-component envelope function solution. The wave function $\psi(x,y)$  can be separated into a product of two factors, one corresponding to plane waves along the $y$ direction and the other corresponding to a nonperiodic function along $x$. This allows us to write the following ansatz \cite{michetti2012helical}%
\begin{equation}\label{ansatzkxmodes}
\psi_{\kappa_x,k_y}(x,y)=\frac{\exp(ik_yy)}{\sqrt{L_y}}%
\begin{pmatrix}
1\\
R_{\kappa_x,k_y}
\end{pmatrix}\exp(i\kappa_xx),
\end{equation}%
where $\kappa_x(k_y, E)$ represents the transversal modes and $R_{\kappa_x,k_y}$ is the ratio between the two wave function components. The normalization constant will be introduced later in the general solution.

Substituting Eq.~\eqref{ansatzkxmodes} into the Schr{\"o}dinger equation, we obtain four modes given by $\kappa_x=(-1)^j\kappa_\pm$, where  $j=\{0,1\}$ and 
\begin{equation}\label{kxmodes}
\kappa_\pm\equiv\sqrt{-k_y^2-F\pm\sqrt{F^2-Q^2}},
\end{equation}%
with
\begin{align}
F\equiv& \frac{A^2-2M B-2D\E-6AGk_y}{2(B^2+G^2-D^2)},\\
Q^2\equiv& \frac{M^2-\E^{2}+8AGk_y^3}{(B^2+G^2-D^2)},\\
\bar{E}\equiv& E-C.
\end{align}%
For each $\kappa_x$, $R_{\kappa_x,k_y}$ is given by%
\begin{equation}\label{Rkxmodes}
R^j_{k_y\pm}=-\frac{iA\left[(-1)^j \kappa_\pm+ik_y\right]+G\left[(-1)^j \kappa_\pm-ik_y\right]^2}{d_{k_y\pm}}, %
\end{equation}
where $d_{k_y\pm}=-M-\bar{E}+(B-D)(\kappa_\pm^2+k_y^2)$. From now on, we use $R^j_{k_y\pm}\equiv R^j_{\pm}$. Using Eqs. (\ref{ansatzkxmodes}), (\ref{kxmodes}), and (\ref{Rkxmodes}), we write the general solution as the linear combination of the four modes: 

\begin{align}
\Psi_{k_y}(x,y)=&\frac{\exp(ik_yy)}{\sqrt{L_y}}\notag\\
&\times\sum_{\substack{j=0,1\\ l=+,-}}c_{j,l}
\begin{pmatrix}
1\\
R^j_l
\end{pmatrix}\exp[i(-1)^j\kappa_l x].\label{gensolution}
\end{align}

By substituting the fitted parameters into Eq. (\ref{kxmodes}) and plotting the imaginary part of $\kappa_{\pm}$, we find $\text{Im}\left(\kappa_{+}\right)>0, \,\, \forall \,\, (k_y,E) \,\in \mathbb{R}^2$. On the other hand, $\text{Im}\left(\kappa_{-}\right)$ can be either positive or negative depending on the values of $(k_y,E)$. For $\text{Im}\left(\kappa_{-}\right)>0$, the general solution, along with the boundary conditions defined in Eqs. (\ref{bondcond1}-\ref{bondcond3}), leads to
\begin{align}
c_{0,-}=&-c_{0,+},\label{const1}\\
c_{1,+}=&c_{1,-}=0,\label{const2}\\
R^0_{+}=&R^0_{-}\label{const3}.
\end{align}
Note that, in applying the boundary conditions in Eqs. (\ref{bondcond1})-(\ref{bondcond3}), we have already neglected the exponential terms that do not result in square-integrable solutions, i.e., that do not yield a vector Hilbert space. Using Eqs. (\ref{gensolution})-(\ref{const3}), we obtain
\begin{align}
\Psi_{k_y}(x,y)=&%
c_{0,+}\frac{\exp(ik_yy)}{\sqrt{L_y}}\notag \\%
&\times\begin{Bmatrix}
\exp(i\kappa_+ x)-\exp(i\kappa_- x)\\
R^0_+\left[\exp(i\kappa_+ x)-\exp(i\kappa_- x)\right]
\end{Bmatrix},%
\end{align}%
where $c_{0,+}$ is determined by normalizing the wave function. Similarly, for $\text{Im}\left(\kappa_{-}\right)<0$, we have%
\begin{align}
c_{1,-}=&-c_{0,+},\label{const1b}\\
c_{1,+}=&c_{0,-}=0,\label{const2b}\\
R^0_{+}=&R^1_{-}\label{const3b},
\end{align}%
and%
\begin{align}
\Psi_{k_y}(x,y)=&%
c_{0,+}\frac{\exp(ik_yy)}{\sqrt{L_y}}\notag \\%
&\times\begin{Bmatrix}
\exp(i\kappa_+ x)-\exp(-i\kappa_- x)\\
R^0_+\left[\exp(i\kappa_+ x)-\exp(-i\kappa_- x)\right]
\end{Bmatrix}.%
\end{align}%

 By solving Eqs. (\ref{const3}) and (\ref{const3b}) for $E$ as a function of $k_y$, we obtain two independent quadratic dispersion relations for the edge states of the spin-up sector. Due to TRS, the two dispersion relations for the edge states with spin-down can be readily calculated by replacing $k_y\rightarrow-k_y$ in the solutions for spin-up. We then have%
\begin{equation}\label{eq:energyexpansion}
E_{i}^{\uparrow(\downarrow)}=b_{i,0}\pm b_{i,1}k_{y}+b_{i,2}k_{y}^{2},\quad (i=1,2),
\end{equation}
where%
\begin{align}
b_{1(2),0}=&C-\frac{DMB\pm\left|MG\right|\sqrt{B^{2}+G^2-D^2}}{B^{2}+G^{2}},\\
b_{1(2),1}=&A\frac{-DG\pm B\,\mathrm{sgn}\left(MG\right)\sqrt{B^{2}+G^2-D^2}}{B^{2}+G^{2}},\\
b_{1(2),2}=&2\frac{-DG^{2}\pm |BG|\sqrt{B^{2}+G^2-D^2}}{B^{2}+G^{2}}\label{eq:b122}
\end{align}%
and $\mathrm{sgn} (x)=+ 1 ~(-1)$ if $x>0$ ($x<0$).

\begin{figure}[h]
    \centering
    \begin{flushleft}\includegraphics[width=\linewidth]{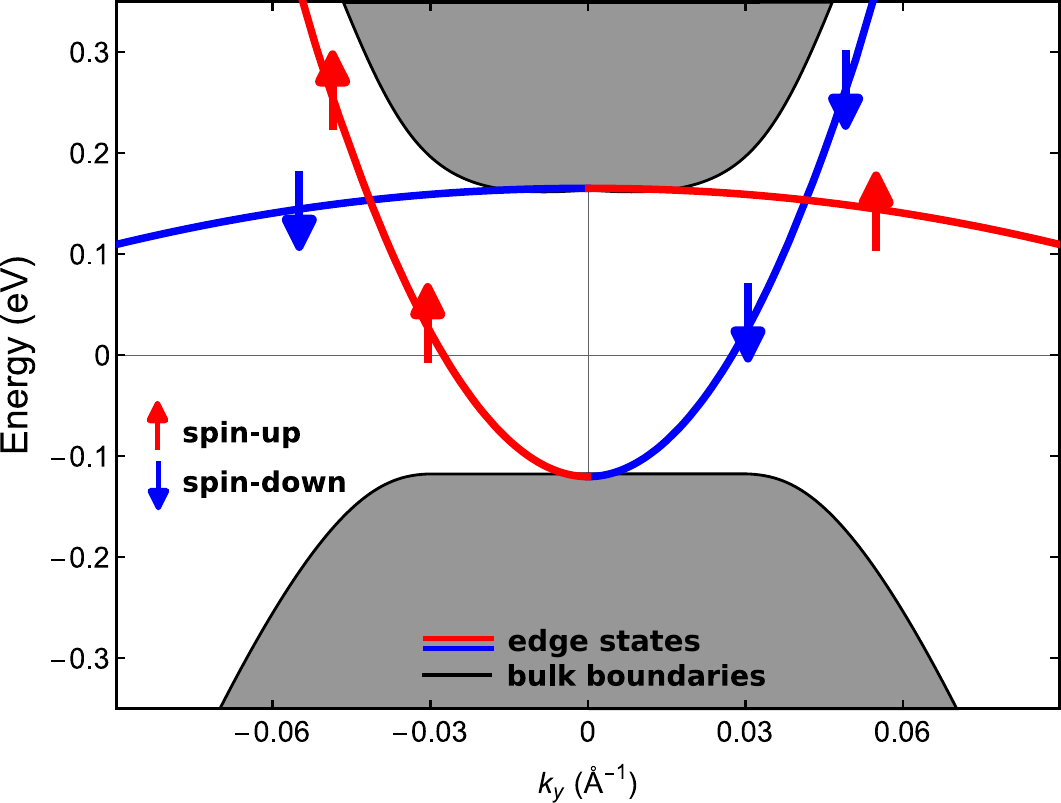}\end{flushleft}
    \caption{(Color online) Topological edge states hosted by the \kp model in a semi-infinite plane geometry, with parameters from the fitting for Na$_2$CdSn. 
    The red (blue) solid lines represent the analytical solutions for the dispersion relations of the spin-up (down) edge states [Eq. (\ref{eq:energyexpansion})], while the solid black lines delimit the bulk region, corresponding to the continua of delocalized bulk states (gray-shaded areas).}
    \label{fig:edgestates_fitted_params}
\end{figure}

In Fig. \ref{fig:edgestates_fitted_params}, we show the analytical dispersion relations of the spin-up (red curve)  and down (blue) edge states [Eq. (\ref{eq:energyexpansion})] for the fitted parameters (Sec. \ref{sec:fitting}). The gray-shaded areas represent the continua of delocalized bulk states (bulk region), whose boundaries are indicated by solid black lines.

For each energy level inside the bulk band gap, there are two pairs of counterpropagating topological edge states of opposite spin. This is in agreement with the bulk-boundary correspondence theorem, considering the topological invariants calculated in Sec. \ref{sec:topologyanalysis}. As it can be seen from Eq. (\ref{eq:b122}), the quadratic dispersion relation of the topological edge states is a consequence of the nonzero  parameter $G$.

Note that all edge states merge into the bulk region at $k_y\approx 0$. As $k_y$ moves away from zero, however, the energy of the edge states increases (in absolute value) slower than the energy of the bulk bands. Hence, here, the edge states never merge into the bulk region again. We emphasize that the effective model derived in Sec. \ref{sec:kpmodel} is only valid in the region close to $k_y=0$; thus, we should not expect it to capture any realistic merging points that might occur at larger values of $k_y$. The absence of merging points is also observed when we consider edge states arising from the BHZ Hamiltonian in the absence of $k^2$ diagonal terms \cite{Candido2018a}.

\subsection{Ribbon geometry}\label{sec:ribbongeometry}

In this section, we apply the finite difference method to obtain the electronic band structure for Na$_2$CdSn in a ribbon (strip) geometry~\cite{JulianZanon2021}. We use the parameters in Table \ref{tab:parametercontributions} and consider open (periodic) boundary conditions along the $x$ ($y$) direction.

\begin{figure*}
    \centering
    \includegraphics[width=.99\linewidth]{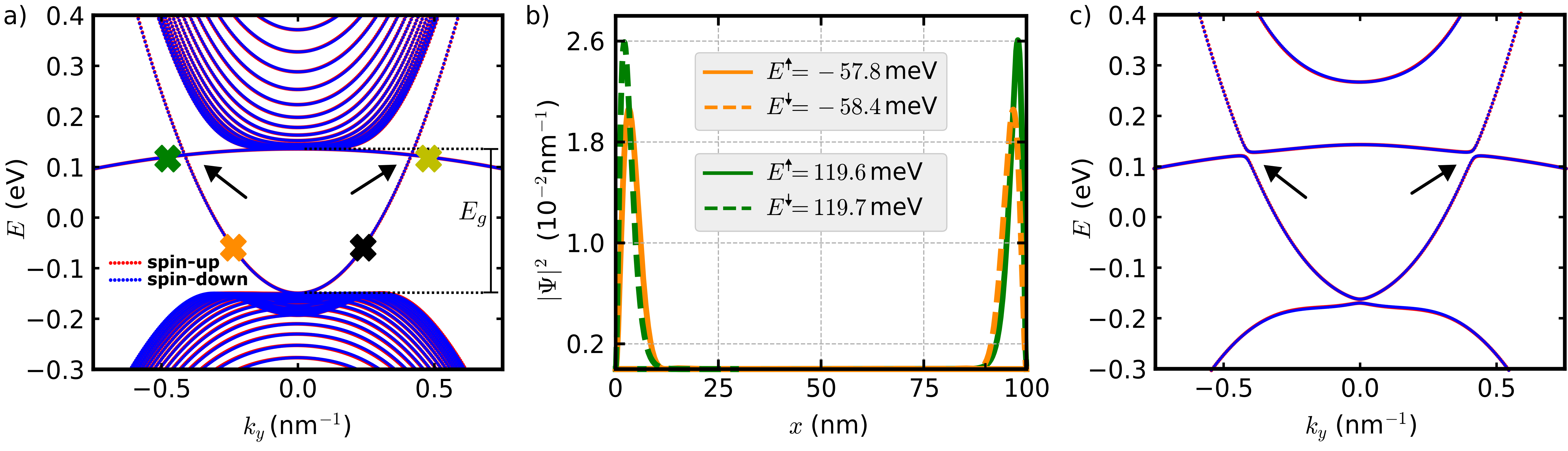}
    \caption{(Color online) Electronic states for Na$_2$CdSn in a ribbon geometry. a) Electronic band structure showing spin-up (red) and spin-down (blue) states with a small spin splitting caused by the bulk inversion asymmetry (BIA) and $\bm{k}$-dependent spin-orbit coupling (SOC). The bulk region is determined by the Mexican-hat-like energy bands, with a bulk band gap of $E_{g} \approx 285$ meV. The quadratic energy bands correspond to quasidegenerate topological edge states, with the spin split states localized at opposite sides of the ribbon. The colored cross marks indicate the selected edge states for our wave function analysis, and the two black arrows denote the points of edge state crossings. b) Probability densities of the selected topological edge states as functions of $x$ (coordinate along the transversal direction). The color of the curves corresponds to the colors of the respective cross marks, while the solid (dashed) style represents spin-up (spin-down) states. c) Electronic band structure for Na$_2$CdSn in a ribbon geometry of width $W=15\,$nm. The black arrows indicate the opening of an anticrossing hybridization gap due to the finite-size effect.}
    \label{fig:ribbonresults}
\end{figure*}

In Fig.~\ref{fig:ribbonresults}-a), we show the electronic band structure for a ribbon of width $W=100\,$nm, i.e., a wide ribbon (compared with the lattice parameter $a=4.978\,$\AA). Each $(k_y,E)$ point is colored according to its spin, with red (blue) representing spin-up (down) states. The energy bands forming a Mexican-hat-like structure constitute the bulk region. The energy gap $E_g\approx 285\,$meV between the lowest bulklike conduction and topmost bulklike valence bands is indicated in the figure. 

The quadratic energy bands crossing the bulk band gap correspond to the topological edge states. All energy bands are quasidegenerate, including the edge states for which the spin splittings are $\sim 10^{-1}$ meV. Like the spin splitting of the energy bands in the infinite bulk system [Fig. \ref{fig:fig2}-c)], the spin splitting of the edge states is not accompanied by a mixing of the spin components due to the block-diagonal structure of the effective Hamiltonian [Eq. (\ref{blockdiagonalhamiltonian})].

The quasidegenerate topological edge states are localized at opposite sides of the strip. Note that, in the wide-ribbon limit, the energy dispersions of the edge states localized on the left (right) side can be mapped onto those of a vacuum/DTI (DTI/vacuum) semi-infinite system. In analogy to Sec. \ref{sec:semiinfiniteplane}, we employ the envelope function approach and solve the problem for the DTI/vacuum geometry. We use the analytical solutions for the two semi-infinite systems to derive an expression for the inter-edge splitting.  Subtracting the dispersion relations for spin-up and spin-down states, we obtained that the spin splittings are given by $2b_{1,1}k_y$ and $2b_{2,1}k_y$ for the concave up and down bands, respectively.

Notice that there is no structural inversion asymmetry in this case, which for the BHZ model is known to generate a Rashba term \cite{liu2008quantum, Rothe2010,krueckl2011switching,krueckl2011switching,ostrovsky2012symmetry,ortiz2016generic} that lifts the degeneracy of the topological edge states \cite{ortiz2016generic}. For the ribbon geometry considered here, the spin splitting arises solely from the BIA. It can be shown that the inclusion of BIA in the BHZ Hamiltonian \cite{michetti2012helical,konig2008the} also leads to interedge spin splitting in 2D TIs, although in this case, the splitting is accompanied by a mixing of the spin components.

In Fig. \ref{fig:ribbonresults}-b), we present the probability density of the edge states, indicated by the green and orange cross marks in Fig. \ref{fig:ribbonresults}-a), as functions of $x$ ($x=0$ and $x=W$ correspond to the ribbon extremities). Each cross mark represents two quasidegenerate states localized at opposite sides of the ribbon  (the line colors match the colors of the corresponding cross mark). The dashed (solid) green line corresponds to a spin-down (spin-up) state localized at the left (right) side of the ribbon. Similarly, the solid (dashed) orange line corresponds to a spin-up (spin-down) state localized at the left (right) side of the ribbon. Note that the states shown by the green color are slightly more localized than the ones represented in orange \footnote{A zoom-in on the probability density distributions shown in Fig. \ref{fig:ribbonresults}-b) reveals that the green (orange) curves present a normal (oscillatory) exponential decay with $x$. The decaying behavior of a given edge state is dictated by the region where it is found in the electronic band structure. This effect is well-discussed for the BHZ model in Ref. \cite{Lu2012}. In our model, however, such an analysis is much more complicated. Therefore, we plan to provide a more detailed investigation of this topic in a forthcoming work.}.

So far, we have only investigated the edge states for negative crystal momentum ($k_y\approx-0.24\,$nm$^{-1}$ and $k_y\approx-0.48\,$nm$^{-1}$). Since our model is invariant under TRS, each topological edge state has a Kramers partner localized on the same side of the ribbon but with opposite momentum and spin. The yellow (black) cross mark in Fig. \ref{fig:ribbonresults}-a) indicates the two Kramers partners of the two states indicated by the green (orange) cross mark. We have verified that, in the wide-ribbon limit, the dispersion relations for the edge states localized at the left side of the ribbon agree exactly with those obtained for the semi-infinite plane geometry [see Eq. (\ref{eq:energyexpansion}) and Fig. \ref{fig:edgestates_fitted_params}].

The edge states indicated by the green cross mark in Fig. \ref{fig:ribbonresults}-a) have a positive group velocity. Therefore, the spin-up state represented by the green solid curve in Fig. \ref{fig:ribbonresults}-b) propagates on the right edge along the $+y$ direction, while the spin-down state represented by the green dashed curve propagates on the left edge, also along the $+y$ direction. Their Kramers partners (yellow cross mark), on the other hand, have a negative group velocity, so there is a spin-down state propagating along the $-y$ direction on the right edge and a spin-up state propagating along the $-y$ direction at the left edge of the ribbon. For the orange cross mark, the edge states have a negative group velocity. Therefore, the states represented by the orange color propagate along the $-y$ direction, while their Kramers partners (black cross mark) propagate along the $+y$ direction.

For any energy level within the bulk band gap (e.g., at the Fermi level), there are four topological edge states at each side of the ribbon. On the left side, there are two spin-down (up) edge states propagating along the $+y$ ($-y$) direction, while on the right side, there are two spin-up (down) edge states propagating along the $+y$ ($-y$) direction.

The black arrows in Fig. \ref{fig:ribbonresults}-a) point to the two edge state crossings taking place around the same energy level $E \approx 0.123\,$eV. At each crossing point, there are four topological edge states. Let us focus on the edge state crossing around $k_{y}\approx - 0.4\,$ nm$^{-1}$ (left black arrow). Two of the four edge states live on the same side of the strip, have opposite spins, and propagate along opposite directions. Since they live in different spin subspaces, they are orthogonal to each other and do not hybridize. The counterpropagating states localized at opposite sides of the ribbon, on the other hand, belong to the same spin subspace, and therefore, can hybridize. Naturally, the hybridization requires a superposition of their wave functions, the so-called finite-size effect \cite{zhou2008finite,Lu2012}. In Fig. \ref{fig:ribbonresults}-c), we show the electronic band structure for Na$_2$CdSn in a ribbon of width $W=15\,$nm. For this narrow ribbon, the edge states (with the same spin) living on opposite sides of the strip hybridize. The hybridization gap due to the anti-crossing of the edge state bands around $k_y\approx-0.4\,$nm$^{-1}$ and $k_{y}\approx 0.4$ nm$ ^{-1}$ are indicated by the two black arrows in the figure.

\section{Conclusions}\label{sec:conclusions}

We have obtained the electronic band structure of the quasi-2D ternary Na$_2$CdSn triatomic layer via DFT calculations and derived an effective $4\times4$ model using the \kp method to describe the low-energy states. To do so, we have employed the Lödwin perturbation theory, the folding-down technique, and the theory of invariants.  Our rigorously derived  effective Hamiltonian contrasts with that of Ref.~\cite{mao2019dual}. It has some quadratic off-diagonal terms (absent in Ref. \cite{mao2019dual}) that are crucial to obtain our spin and mirror Chern numbers and the corresponding (two pairs of) protected edge states.

More specifically, by  fitting our effective model to the DFT energy bands, we have obtained numerical values for the model parameters and predicted that Na$_2$CdSn is a DTI with spin and mirror Chern numbers given by $C_s=-2$ and $C_m=-2$, respectively. Using the envelope function approach, we have also obtained analytical dispersion relations, quadratic in the crystal momentum, for the topological edge states in the semi-infinite plane geometry. The edge modes have also been analyzed numerically in the ribbon geometry. For both geometries, we have found two pairs of counterpropagating topological edge states at each boundary, as expected from the bulk-boundary correspondence.  Both our Chern numbers and the corresponding (number of) edge states are in blatant contrast with the results in Ref. \cite{mao2019dual}; there, spin and mirror Chern numbers $-1$ were reported and, consequently, only one pair of topological edge states.

Our investigation contributes to the understanding of Na$_2$CdSn as a giant gap ($E_g = 234.8\,$meV) DTI in 2D by providing its proper model Hamiltonian and the corresponding effective $\bm{k}\cdot\bm{p}$ parameters. This enables us to derive the energy dispersions and wave functions of the edge states arising in finite geometries of this material. The Na$_2XY$ ($X$= Mg, Cd; $Y$= Pb, Sn) family of triatomic layers are the  building stacking blocks of 3D van der Waals Dirac semimetals, already synthesized experimentally \cite{schuster1980ternare,peng2018}. They can also be  potentially obtained via mechanical exfoliation. The DTI Na$_2$CdSn represents a promising alternative to conventional small-gap TIs in 2D -- which usually require the engineering of elaborated quantum wells \cite{bernevig2006quantum,konig2007quantum,liu2008quantum,Liu2008a,Candido2018}. We are optimistic that our findings will stimulate the experimental verification of the DTI phase in this outstanding candidate for realistic room-temperature applications in nanotechnology, such as nanoelectronic, spintronic, thermoelectric, and optical devices. As more specific possibilities, we envisage the implementation of Na$_2$CdSn in topological transistors \cite{Xu2019,acosta2019spin} and topological quantum point contacts \cite{strunz2020interacting} for increased device robustness.

\section{Acknowledgments}\label{sec:acknowledgments}

W.H.C. acknowledges helpful discussions with Michael Flatt{\'e} and Jaroslav Fabian. This paper was supported by the National Council for Scientific and Technological Development (CNPq) Grant No. 301595/2022-4, PDJ Process No. 152321/2020-9, and Grant No. 306122/2018-9; by Coordena{\c c}{\~a}o de Aperfei{\c c}oamento de Pessoal de N{\' i}vel Superior (PNPD/CAPES); by S{\~a}o Paulo Research Foundation (FAPESP), Grant No. 2020/00841-9. J.Z. acknowledges the financial support of the Quantimony (Marie Sklodowska-Curie Project Grant No. 956548). P.E.F.J and W.H.C. acknowledge the financial support of the Deutsche Forschungsgemeinschaft (DFG, German Research Foundation) SFB 1277 (Project-ID No. 314695032, Projects No. B07 and No. B11). P.E.F.J. acknowledges the financial support of the DFG-SPP 2244 (Project No. 443416183). W.H.C. acknowledges funding from the Deutsche Forschungsgemeinschaft (DFG) Grant No. TRR 173 268565370 (Project A03).

\appendix%

\section{Mirror symmetry operation}\label{sec:mirrorsymmetry}

To properly define the mirror symmetry operation about the $z$ axis (in this case, the mirror plane is the $xy$ plane), we consider its effect on both spin and spatial degrees of freedom since spinful electrons live in the state space given by the direct product between the spin and  orbital subspaces \cite{cohen2019quantum,sakurai2017modern}. The spin angular momentum transforms as an axial vector, and therefore, the components of the spin operator perpendicular and parallel to the mirror plane must satisfy, respectively, %
\begin{equation}
\left[\hat{\sigma}_h,\hat{S}_z\right]=0\quad \text{and} \quad \left\{\hat{\sigma}_h,\hat{S}_{i}\right\}=0,\quad \label{eq:Mzparall}
\end{equation}%
with $i=x,y$~\cite{chiu2016classification}. The mirror symmetry operator,  consistent with Eq. (\ref{eq:Mzparall}), is given by the direct product:%
\begin{equation}\label{eq:mirroropstatespace}
    \hat{\sigma}_h=\hat{\sigma}_h^{spin}\otimes \hat{\sigma}_h^{orb},
\end{equation}%
where $\hat{\sigma}_h^{spin}$  and $\hat{\sigma}_h^{orb}$ are the mirror symmetry operators in the spin and orbital vector subspaces, respectively.

The $\hat{\sigma}_h^{spin}$ operator can be defined as a $\pi$ rotation about the $z$ axis followed by the inversion operation (both in the spin subspace) \cite{teo2008surface,Ando2015,Gao2016}, that is,

\begin{equation}
    \hat{\sigma}_h^{spin}=\hat{I}^{spin}\hat{R}_z^{spin}(\pi).
\end{equation}%
The inversion operator does not affect spin degrees of freedom, and therefore, it corresponds to the identity operator in the spin subspace \cite{teo2008surface}. As usual, the rotation operator is given by \cite{sakurai2017modern}%
\begin{equation}
    \hat{R}_z^{spin}(\pi)=\exp\left(-i\frac{\hat{S}_z}{\hbar}\pi\right)=-i\hat{\sigma}_z,
\end{equation}%
where we have used $\hat{S}_z=\frac{\hbar}{2}\hat{\sigma}_z$. Hence, the mirror operator in the state space [Eq. (\ref{eq:mirroropstatespace})] reads%
\begin{equation}\label{eq:mirrorstatespace}
    \hat{\sigma}_h=-i\hat{\sigma}_z\otimes \hat{\sigma}_h^{orb}.
\end{equation}%

Applying the $\hat{\sigma}_h^{orb}$ operator twice returns the spatial functions to their original state, so that $\left(\hat{\sigma}_h^{orb}\right)^2=1$. However, when spin is considered, applying the mirror operation twice leads to a change of sign
\begin{equation}
\hat{\sigma}_h^{2}=-1,
\end{equation}
which yields the eigenvalues of $\hat{\sigma}_h$ being $i$ and $-i$ \cite{teo2008surface,Ando2015}.

To obtain the matrix representation of $\hat{\sigma}_h$, we can use Eq. (\ref{eq:mirrorstatespace}), and project $\hat{\sigma}_z$ and $\hat{\sigma}_h^{orb}$ onto the basis of the spin and orbital subspaces, respectively. The basis functions for the orbital subspace are determined by the irreps of the simple group relevant for the material under consideration.

The mirror symmetry has important consequences for the spin texture of the energy bands. Let $|\psi_{\bm{k},\eta}\rangle$ be a Bloch state with mirror eigenvalue $\eta$. Taking the expectation values of the spin components gives%
\begin{equation}
\begin{split}
    \left\langle  \hat{S}_z \right\rangle =&  \left\langle \psi_{\bm{k},\eta} \middle| \hat{\sigma}_h^{-1}\hat{\sigma}_h \hat{S}_z \hat{\sigma}_h^{-1}\hat{\sigma}_h \middle| \psi_{\bm{k},\eta} \right\rangle\\
    =& |\eta|^2 \left\langle  \hat{S}_z  \right\rangle=\left\langle  \hat{S}_z\right\rangle
\end{split}
\end{equation}%
and%

\begin{equation}
\begin{split}
    \left\langle  \hat{S}_{i} \right\rangle =&  \left\langle \psi_{\bm{k},\eta} \middle| \hat{\sigma}_h^{-1}\hat{\sigma}_h \hat{S}_{i} \hat{\sigma}_h^{-1}\hat{\sigma}_h \middle| \psi_{\bm{k},\eta} \right\rangle  = - |\eta|^2 \left\langle  \hat{S}_{i} \right\rangle\\
    =& -\left\langle  \hat{S}_{i} \right\rangle=0 \quad (i=x,y),
\end{split}
\end{equation}%
where we have used Eq. (\ref{eq:Mzparall}). Therefore, for the $|\psi_{\bm{k},\eta}\rangle$ Bloch state the spin texture must be polarized along the $z$ direction in the whole BZ, except at degeneracies \cite{kurpas2019spin}.\\

\section{Expressions for the model parameters in the L{\"o}wdin perturbation approach}\label{sec:expressionsparametersLowdin}%

In Sec. \ref{subsec:lowdinresults}, we have shown the effective \kp Hamiltonian derived using the L{\"o}wdin perturbation theory. Below, we show the explicit expressions for the parameters of the model. The parameters $A_1$ and $A_2$ are given, respectively, by

\begin{widetext}
	
	\begin{equation}
		A_1=\left(\frac{\hbar}{m_0}\right)^2\left[\sum_\beta^{B[\Gamma_1]}\frac{|\langle X|p_x|\Gamma_{1\beta}\rangle|^2}{\epsilon_0-E_{\Gamma_{1\beta}}}+\sum_\beta^{B[\Gamma_6]}\frac{|\langle X|p_x|\Gamma_{6\beta}^x\rangle|^2}{\epsilon_0-E_{\Gamma_{6\beta}}}\right], \label{eq:paramA1}
	\end{equation}
and
	\begin{equation}
		A_2=\left(\frac{\hbar}{m_0}\right)^2\left[\sum_\beta^{B[\Gamma_2]}\frac{|\langle X|p_y|\Gamma_{2\beta}\rangle|^2}{\epsilon_0-E_{\Gamma_{2\beta}}}+\sum_\beta^{B[\Gamma_6]}\frac{|\langle X|p_x|\Gamma_{6\beta}^x\rangle|^2}{\epsilon_0-E_{\Gamma_{6\beta}}}\right],
	\end{equation}%
\end{widetext}%
where we have used $\langle X|p_y|\Gamma_{6\beta}^y\rangle=-\langle X|p_x|\Gamma_{6\beta}^x\rangle$  (similar relations were used to simplify all the matrix elements below). In the expressions above, $B[\Gamma_i]$ indicates that $\beta$ runs over all infinite remote bands belonging to the $\Gamma_i$ irrep. When the irrep of the $\beta$th band is 1D, its single basis function is written as $|\Gamma_{i\beta}\rangle$. For irreps of higher dimensions, we use $|\Gamma_{i\beta}^\mu\rangle$ to denote the basis function that transforms according to $\mu$, i.e., $|\Gamma_{i\beta}^\mu\rangle \sim \mu$, $(\mu=x,y,R_x,R_y...)$. Here, $E_{\Gamma_{i\beta}}$ is the energy of the $\beta$th band belonging to the $\Gamma_i$ irrep. Naturally, all basis functions of an irrep have the same energy eigenvalue, and therefore, it is not necessary to label the energy eigenvalue with the basis function. The remaining parameters read%
\begin{widetext}
	\begin{equation}
		D_1=-\frac{\hbar^2}{4m_0^3c^2}\left[2\sum^{B[\Gamma_6]}_\beta \frac{\langle X |p_x|\Gamma_{6\beta}^x\rangle\langle\Gamma_{6\beta}^y|A_z|X\rangle}{\epsilon_0-E_{\Gamma_{6\beta}}}\right],
	\end{equation}%
	\begin{equation}\label{eq:expressionforBparameter}
			B=-i\frac{\hbar^3}{4m_0^3c^2}\Bigg[\sum^{B[\Gamma_1]}_\beta\frac{\langle X|p_x|\Gamma_{1\beta}\rangle\langle \Gamma_{1\beta}|\partial_xV|X\rangle}{\epsilon_0-E_{\Gamma_{1\beta}}}+\sum^{B[\Gamma_2]}_\beta\frac{\langle X|p_y|\Gamma_{2\beta}\rangle\langle \Gamma_{2\beta}|\partial_yV|X\rangle}{\epsilon_0-E_{\Gamma_{2\beta}}}-2\sum^{B[\Gamma_6]}_\beta\frac{\langle X|p_x|\Gamma_{6\beta}^x\rangle\langle \Gamma_{6\beta}^x|\partial_xV|X\rangle}{\epsilon_0-E_{\Gamma_{6\beta}}}\Bigg],
	\end{equation}%
	\begin{equation}
		C_1=\left(\frac{\hbar}{4m_0^2c^2}\right)^2\left[%
		\sum^{B[\Gamma_6]}_\beta\frac{|\langle X|A_z|\Gamma_{6\beta}^y\rangle|^2}{\epsilon_0-E_{\Gamma_{6\beta}}}+\sum^{B[\Gamma_3]}_\beta\frac{|\langle X|A_x|\Gamma_{3\beta}\rangle|^2}{\epsilon_0-E_{\Gamma_{3\beta}}} +\sum^{B[\Gamma_4]}_\beta\frac{|\langle X|A_y|\Gamma_{4\beta}\rangle|^2}{\epsilon_0-E_{\Gamma_{4\beta}}}+2\sum^{B[\Gamma_5]}_\beta\frac{|\langle X|A_x|\Gamma_{5\beta}^{R_x}\rangle|^2}{\epsilon_0-E_{\Gamma_{5\beta}}}%
		\right],
	\end{equation}
	\begin{equation}
		C_2=\left(\frac{\hbar}{4m_0^2c^2}\right)^2\left[\sum^{B[\Gamma_3]}_\beta\frac{|\langle X|A_x|\Gamma_{3\beta}\rangle|^2}{\epsilon_0-E_{\Gamma_{3\beta}}}+\sum^{B[\Gamma_4]}_\beta\frac{|\langle X|A_y|\Gamma_{4\beta}\rangle|^2}{\epsilon_0-E_{\Gamma_{4\beta}}}-2\sum^{B[\Gamma_5]}_\beta\frac{|\langle X|A_x|\Gamma_{5\beta}^{R_x}\rangle|^2}{\epsilon_0-E_{\Gamma_{5\beta}}}\right],
	\end{equation}%
	\begin{equation}
		E_1=-i\frac{\hbar^3}{(4m_0^2c^2)^2}\left[2\sum^{B[\Gamma_5]}_\beta\frac{\langle X|A_x|\Gamma_{5\beta}^{R_x}\rangle\langle\Gamma_{5\beta}^{R_y}|\partial_zV|X\rangle}{\epsilon_0-E_{\Gamma_{5\beta}}}\right],
	\end{equation}%
	\begin{equation}
		F_1=\left(\frac{\hbar^2}{4m_0^2c^2}\right)^2\left[\sum^{B[\Gamma'_2]}_\beta\frac{|\langle X|\partial_yV|\Gamma_{2\beta}\rangle|^2}{\epsilon_0-E_{\Gamma_{2\beta}}}+\sum^{B[\Gamma_6]}_\beta\frac{|\langle X|\partial_xV|\Gamma_{6\beta}^x\rangle|^2}{\epsilon_0-E_{\Gamma_{6\beta}}}+\sum^{B[\Gamma_5]}_\beta\frac{|\langle X|\partial_zV|\Gamma_{5\beta}^{R_y}\rangle|^2}{\epsilon_0-E_{\Gamma_{5\beta}}}\right],
	\end{equation}%
	and
	\begin{equation}\label{eq:paramF2}
		F_2=\left(\frac{\hbar^2}{4m_0^2c^2}\right)^2\left[\sum^{B[\Gamma_1]}_\beta\frac{|\langle X|\partial_xV|\Gamma_{1\beta}\rangle|^2}{\epsilon_0-E_{\Gamma_{1\beta}}}+\sum^{B[\Gamma_6]}_\beta\frac{|\langle X|\partial_xV|\Gamma_{6\beta}^x\rangle|^2}{\epsilon_0-E_{\Gamma_{6\beta}}}+\sum^{B[\Gamma_5]}_\beta\frac{|\langle X|\partial_zV|\Gamma_{5\beta}^{R_y}\rangle|^2}{\epsilon_0-E_{\Gamma_{5\beta}}}\right].
	\end{equation}%
\end{widetext}

\section{Model derivation via the Theory of Invariants}\label{sec:theoryofinvariants}

As discussed in the main text, we have also derived the $4 \times 4$ effective model in Eqs~\eqref{eqH0}--\eqref{eq:2ndorder2} by using the theory of invariants~\cite{winkler2003book, bir1974book}. This approach allows one to obtain a finite-dimensional \kp Hamiltonian $H(\bm{k})$ based only on the symmetries of the crystal under consideration. More specifically, it consists of writing $H(\bm{k})$ as a Taylor expansion in the momentum $\bm{k}$,
\begin{equation}
H(\bm{k}) = \sum_{l,m,n} h_{l,m,n} k_x^l k_y^m k_z^{n},
\label{eq:expansion}
\end{equation}
and finding the unknown coefficients $h_{l,m,n}$ (matrices) by imposing that the Hamiltonian $\mathcal{H}$ of the crystal, characterized by a point group $G$,  be invariant under all symmetry operations $g_i$ ($g_i \in G$), i.e., $[\mathcal{H}, g_i]=0$. For the Bloch Hamiltonian $\mathcal{H}(\bm{k})=\exp(-i\bm{k}\cdot\bm{r})\mathcal{H}\exp(i\bm{k}\cdot\bm{r})$, this commutator yields
\begin{equation}
\mathcal{H}(u_{g_{i}}\bm{k}) = g_i \mathcal{H}(\bm{k}) g_i^{-1},
\end{equation}
where $u_{g_{i}}$ is the matrix representation of $g_i$ in $\bm{k}$ space. An equivalent expression can be derived for the projected Hamiltonian $H(\bm{k})$ in Eq.~\eqref{eq:expansion}, given by
\begin{equation}
H(u_{g_{i}}\bm{k}) = U_{g_i} H(\bm{k}) U_{g_i}^{-1},
\label{eq:gt}
\end{equation}
with $U_{g_i}$ the matrix representation of $g_i$ in the Hilbert space.
Hence, by using Eq.~\eqref{eq:expansion} and the constraints imposed by Eq.~\eqref{eq:gt}, we obtain a system of equations for the coefficients $h_{l,m,n}$. Note that, for 2D systems, which is the case we are interested in here,
we can simply drop $k_z$ and the sum over $n$ in Eq.~\eqref{eq:expansion}.

The Na$_2$CdSn triatomic layer we are interested in belongs to the point group $D_{3h}$. To determine $h_{l,m,n}$, it is sufficient to use in Eq.~\eqref{eq:gt} only the corresponding group generators $\{g_i\} = \{C_{3z},C_{2x}, \sigma_h\}$, with $C_{3z}$ representing a rotation about the $z$ axis, $C_{2x}$ a rotation about the $x$ axis, and $\sigma_h$ a mirror plane reflection about the $xy$ plane (see character table in the SM~\cite{Na2CdSnsupmat}). In addition, we consider that the system is time-reversal symmetric. 

Here, we use the QSYMM Python package~\cite{varjas2018qsymm}, which receives as input the matrix representations $U_{g_{i}}$ and ${g_i}$, for all $g_i \in G$ (generators + TRS), and returns the matrices $h_{l,m,n}$. The matrix representations $U_{g_i}$ of the generators $\{C_{3z},C_{2x}, \sigma_h\}$ and time-reversal operator $\mathcal{T}$ follow from both the 2D irrep of the orbital part ($\Gamma_6$) and the spinorial part ($\Gamma_7$), i.e.,  $\Gamma_6 \otimes \Gamma_7 = \Gamma_8 \oplus \Gamma_9$, and are given by
\begin{eqnarray}
U_{C_{3z}} &=&
    \exp\left(-\frac{i\phi\sigma_3}{2}\right) \otimes \exp(-i\phi\tau_2), \\
U_{C_{2x}} &=& -i\sigma_1 \otimes \tau_3, \\
U_{\sigma_{h}} &=& -i\sigma_3 \otimes \tau_0, \\
T &=& -i\sigma_2 \otimes \tau_0 K,
\end{eqnarray}
with $\phi=2\pi/3$.
Up to second order in the crystal momentum $\bm{k}$, the $4\times 4$ Hamiltonian is given by

\begin{widetext}
\begin{eqnarray}
H(\bm{k}) &=& \begin{pmatrix}
c_0 & ic_1 & 0 & 0 \\
-ic_1 & c_0 & 0 & 0 \\
0 & 0 & c_0 & -ic_1 \\
0 & 0 & ic_1 & c_0 
\end{pmatrix} + \begin{pmatrix}
c_3 k_y & c_3 k_x & 0 & 0 \\
c_3 k_x& - c_3 k_y& 0 & 0 \\
0 & 0 & - c_3 k_y & - c_3 k_x \\
0 & 0 & - c_3 k_x& c_3 k_y
\end{pmatrix} \nonumber \\
&+& \begin{pmatrix}
c_4 k_y^2 + c_5 k_x^2 & (c_5 - c_4) k_xk_y + i c_6 k^2 & 0 & 0 \\
(c_5 - c_4) k_xk_y  - i c_6 k^2 & c_4 k_x^2 + c_5 k_x^2 & 0 & 0 \\
0 & 0 & c_4 k_y^2 + c_5 k_x^2 & (c_5 - c_4) k_xk_y  - i c_6 k^2 \\
0 & 0 & (c_5 - c_4) k_xk_y  + i c_6 k^2 & c_4 k_x^2 + c_5 k_y^2
\end{pmatrix},
\label{eq:inv}
\end{eqnarray}
\end{widetext}
with $k^2=k_x^2+k_y^2$.

Equation~\eqref{eq:inv} is the most general form of the Hamiltonian $H(\bm{k})$ allowed by symmetry for our 2D system. A detailed investigation using the results in Sec.~\ref{sec:kpmodel} (see also the SM~\cite{Na2CdSnsupmat}) yields
\begin{eqnarray}
	c_0 &=& \epsilon_0 + C_1, \\
	c_1 &=& -\alpha - C_2, \\
	c_3 &=& \zeta + D_1 + E_1, \\
	c_4 &=& A_2 + F_2, \\
	c_5 &=& A_1 + F_1, \\
	c_6 & = -B.
\end{eqnarray}

\bibliography{library}

\end{document}